\documentclass{statsoc}
\usepackage{epsf} %% for .ps file
\usepackage{amssymb,amsmath}
\usepackage{mathrsfs}
\usepackage{amsfonts}
\usepackage{graphicx}
\usepackage{natbib}
\usepackage{verbatim}
\usepackage{color}

\newtheorem{thm}{Theorem}[section]

\title[The DECME Algorithm]{The Dynamic ECME Algorithm}
\author[Y. He {\it et al.}]{Yunxiao He}
\address{Yale University, New Haven, USA}
\email{yunxiao.he@yale.edu}
\author{Chuanhai Liu}
\address{Purdue University, West Lafayette, USA}

\begin{document}

\begin{abstract}
The ECME algorithm has proven to be an effective way of accelerating
the EM algorithm for many problems. Recognising the limitation of
using prefixed acceleration subspaces in ECME, we propose a new
Dynamic ECME (DECME) algorithm which allows the acceleration
subspaces to be chosen dynamically. Our investigation of the
classical Successive Overrelaxation (SOR) method, which can be considered as a special case of DECME, leads to an efficient, simple, stable, and widely applicable DECME implementation, called DECME\_v1. The fast convergence of DECME\_v1 is established by
the theoretical result that, in a small neighbourhood of the maximum
likelihood estimate (MLE), DECME\_v1 is equivalent to a conjugate
direction method. Numerical results show that DECME\_v1 and its two
variants often converge faster than EM by a factor of one hundred in
terms of number of iterations and a factor of thirty in terms of CPU
time when EM is very slow. 
\end{abstract}

%%%%%%%%%%%%%%%%%%%%%%%%%%%%%%%%%%%%%%%%%%%%%%%%%%%%%%%%%%%%%%%%%%%%%%%%%%%%%%%%%%%%
\section{Introduction} \label{sec:introduction}

After its booming popularity of 30 years since the publication of
\citet{DLR:1977}, the EM algorithm is still expanding its
application scope in various areas. At the same time, to overcome
the slow convergence of EM, quite a few extensions of EM have been
developed in such a way that they run faster than EM while
maintaining its widely recognised simplicity and stability. We refer
to \citet{Varadhan:Roland:2008} for a recent nice review of various
methods for accelerating EM. In the present paper, we start by
exploring the convergence of the ECME algorithm
\citep{Liu:Rubin:ecme:1994}, which has proved to be a simple and
effective method  to accelerate its parent EM algorithm \citep[see,
{\it e.g.},][]{sammel:Ryan:1996, kowalski:Tu:Day:1997,
Pinheiro:Liu:Wu:2001}, to name a few.

ECME is a simple extension of the ECM algorithm
\citep{Meng:Rubin:ECM:1993} which itself is an extension of EM.
These three algorithms are summarised as follows. Let $Y_{obs}$ be
the observed data. Denote by $L(\theta|Y_{obs}),\ \ \theta \in
\Theta \subset \mathcal {R}^{p}$, the observed log-likelihood
function of $\theta$. The problem is to find the MLE $\hat{\theta}$ that
maximises $L(\theta|Y_{obs})$. Let $Y=(Y_{obs}, \ Y_{mis})$
represent the complete data with $Y_{obs}$ augmented by the missing
data $Y_{mis}$.  As an iterative algorithm, the $t$th iteration of
EM consists of the E-step, which computes
$Q(\theta|Y_{obs},\theta_{t-1})$, the expected complete-data
log-likelihood function given the observed data and the current
estimate $\theta_{t-1}$ of $\theta$, and the M-step, which finds
$\theta=\theta_{t}$ to maximise $Q(\theta|Y_{obs},\theta_{t-1})$.

 The ECM algorithm replaces the M-step with a sequence of simpler constrained or conditional
maximisation (CM) steps, indexed by $s = 1,\cdots,S$, each of which
fixes some function of $\theta$, $h_s(\theta)$. The ECME algorithm
further partitions the $S$ CM-steps into two groups
$\mathscr{S}_{Q}$ and $\mathscr{S}_{L}$ with $\mathscr{S}_{Q} \cup
\mathscr{S}_{L}=\{1, \cdots, S\}$. While the CM-steps indexed by $s
\in \mathscr{S}_{Q}$ (refereed to as the $MQ$-steps) remain the same
with ECM, the CM-steps indexed by $s \in \mathscr{S}_{L}$ (refereed
to as the $ML$-steps) maximise $L(\theta|Y_{obs})$ in the subspace
induced by $h_s(\theta)$. A more general framework that includes ECM
and ECME as special cases is developed in \citet{Meng:van:AEM:1997}.
However, most of the practical algorithms developed under this
umbrella belong to the scope of a simple case, {\it i.e.,} the
parameter constraints are formed by creating a partition,
$\mathscr{P}$, of $\theta$ as $(\theta_1, \ \cdots, \ \theta_S)$
with associated dimensions $(d_1, \ \cdots, \ d_S)$. Mathematically
we have $h_s(\theta)=(\theta_1, \ \cdots, \ \theta_{s-1}, \
\theta_{s+1}, \cdots, \ \theta_S)$ for $s=1, \cdots, S$.

The advantage of ECME over EM in terms of efficiency depends on the
relationship between the slowest converging directions
 of EM and the acceleration subspaces of ECME, {\it i.e.}, the subspaces for the $ML$-steps. For
example, when the former is effectively embedded within the latter,
ECME achieves its superior gain of efficiency over its parent EM. In
practise, we usually have no information about the convergence of EM
before obtaining the MLE and cannot select the prefixed
acceleration subspaces of ECME accordingly. Hence small or minor
efficiency gain by ECME is expected in some situations. This is
illustrated by the two examples in Section 2 and motivates the idea
of dynamically constructing subspaces for applying the $ML$-step.
This idea is formulated as the generic DECME algorithm. It includes
SOR as a special case. SOR was first developed as an accelerator for
a class of iterative solvers of linear systems in 1950's
\citep{frankel:1950, young:1954}. The same idea has been frequently
explored in the context of EM \citep[among many others although
sometimes under different names]{Salakhutdinov03adaptiveoverrelaxed,
Hesterberg:stag:2005}. However, as shown later, SOR suffers from
what is known as the zigzagging problem. Hence it is often
inefficient.

Motivated by the zigzagging phenomenon observed on SOR, we propose
an efficient DECME implementation, called DECME\_v1.
It is shown that, under some common assumptions, DECME\_v1
is equivalent to a conjugate direction method, which has been proposed in
several different contexts, {\it e.g.,} solving linear systems
\citep{Concus:Golub:O'Leary:1976} and nonorthogonal analysis of
variance \citep{Golub:Nash:1982}.
\cite{Jamshidian:1993} propose to use the conjugate direction method
to accelerate EM. They call the resulting method AEM and
demonstrate its dramatically improved efficiency.
However, AEM is not as popular as one would expect it to be.
This is perhaps due to its demands for extra efforts for coding the
gradient vector of $L(\theta|Y_{obs})$, which is problem specific
and can be expensive to evaluate.

Compared to AEM, DECME\_v1 is simpler to implement because it does
not require computing the gradient of $L(\theta|Y_{obs})$. It does
require function evaluations, which are typically coded with EM
implementation for debugging and monitoring convergence. As SOR, the
only extra requirement for implementing DECME\_v1 is a simple line
search scheme. Such a line search scheme can be used for almost all
EM algorithms for different models. To reduce the number of function
evaluations, two variants of DECME\_v1, called DECME\_v2 and
DECME\_v3, are also considered. Numerical results show that all the
three new DECME implementations obtain dramatic efficiency
improvement over EM, ECME, and SOR in terms of both number of
iterations and CPU time.

The remaining of the paper is arranged as follows. Section
\ref{sec:ecme} provides a pair of motivating ECME examples. Section
\ref{sec:decme_sor} defines the generic DECME algorithm, discusses
the convergence of SOR, and proposes the three efficient novel
implementations of DECME. Section \ref{sec:numerical} presents
several numerical examples to
 compare the performance of different methods.
Section \ref{sec:discussion} concludes with a few remarks.

\section{Two Motivating ECME Examples}\label{sec:ecme}
Following \citet{DLR:1977}, in a small neighbourhood of
$\hat{\theta}$, we have approximately
\begin{equation}
\label{equ:EM}
 \hat{\theta}-\theta_{t}= DM^{EM}(\hat{\theta}-\theta_{t-1}),
\end{equation}
where the $p\times p$ matrix $DM^{EM}$ is known as the missing
information fraction and determines the convergence rate of EM. More
specifically, each eigenvalue of $DM^{EM}$ determines the
convergence rate of EM along the direction of its corresponding
eigenvector (see review in Appendix \ref{app:EM}).

It is shown in \citet{Liu:Rubin:ecme:1994} that ECME also has a
linear convergence rate determined by the $p \times p$ matrix
$DM^{ECME}$ that plays the same role for ECME as $DM^{EM}$ does for
EM. Obviously, ECME will be faster than EM if the largest eigenvalue
of $DM^{ECME}$ is smaller than that of $DM^{EM}$. With the following
two examples we illustrate that it is the choice of the acceleration
subspaces by ECME that determines the relative magnitude of the
dominating eigenvalues of $DM^{EM}$ and $DM^{ECME}$, and hence the
relative efficiency of EM and ECME. All the numerical examples in
this paper are implemented in R \citep{R}.

\subsection{A Linear Mixed-effects Model Example} \label{sec:rat}
Consider the rat population growth data in \citet[Tables 3,
4]{Gelfand:Hills:1990}. Sixty young rats were assigned to a control
group and a treatment group with $n=30$ rats in each. The weight of
each rat was measured at ages $x = 8,\ 15,\ 22,\ 29$ and $36$ days.
We denote by $y_{i}^{g}$ the weights of the $i$th rat in group $g$
with $g = c$ for the control group and $g = t$ for the treatment
group. The following linear mixed-effects model
\citep{Laird:Ware:1982} is considered in \citet{Liu:infoM:1998}:
\begin{equation}
y_{i}^{g}| \theta \sim N(X \beta_g+ X b_i^{g}, \ \sigma_g^2 I_5), \
\ b_i^{g} \sim N(0, \Psi),
\end{equation}
 for $i = 1, \cdots, n$ and $g = c$ and $t$, where $X$ is the $5 \times 2$ design matrix with a
vector of ones as its first column and the vector of the five
age-points as its second column, $\beta_g =
(\beta_{g,1},\beta_{g,2})'$ contains the fixed effects, $b_i^{g}
=(b_{i,1}^{(g)},b_{i,2}^{(g)})'$ contains the random effects, $\Psi
> 0$ is the $2 \times 2$ covariance matrix of the random effects,
and $\theta$ is the vector of the parameters, that is,
$\theta=(\beta_{c,1},\beta_{c,2}, \beta_{t,1},\beta_{t,2},
\Psi_{1,1},\Psi_{1,2},\Psi_{2,2}, \sigma_c^2, \sigma_t^2)'$. Let
$\beta=(\beta_{c,1},\beta_{c,2}, \beta_{t,1},\beta_{t,2})'$ and
$\sigma^2=(\sigma_c^2, \sigma_t^2)'$. The starting point for running EM and ECME is chosen
to be $\beta=(0,0,0,0)',\ \sigma^2=(1,1)'$, and $\Psi=I_{2}$. The
stopping criterion used here is given in Section \ref{sec:setting}.

\begin{figure}
 \centering
\includegraphics[width=3in]{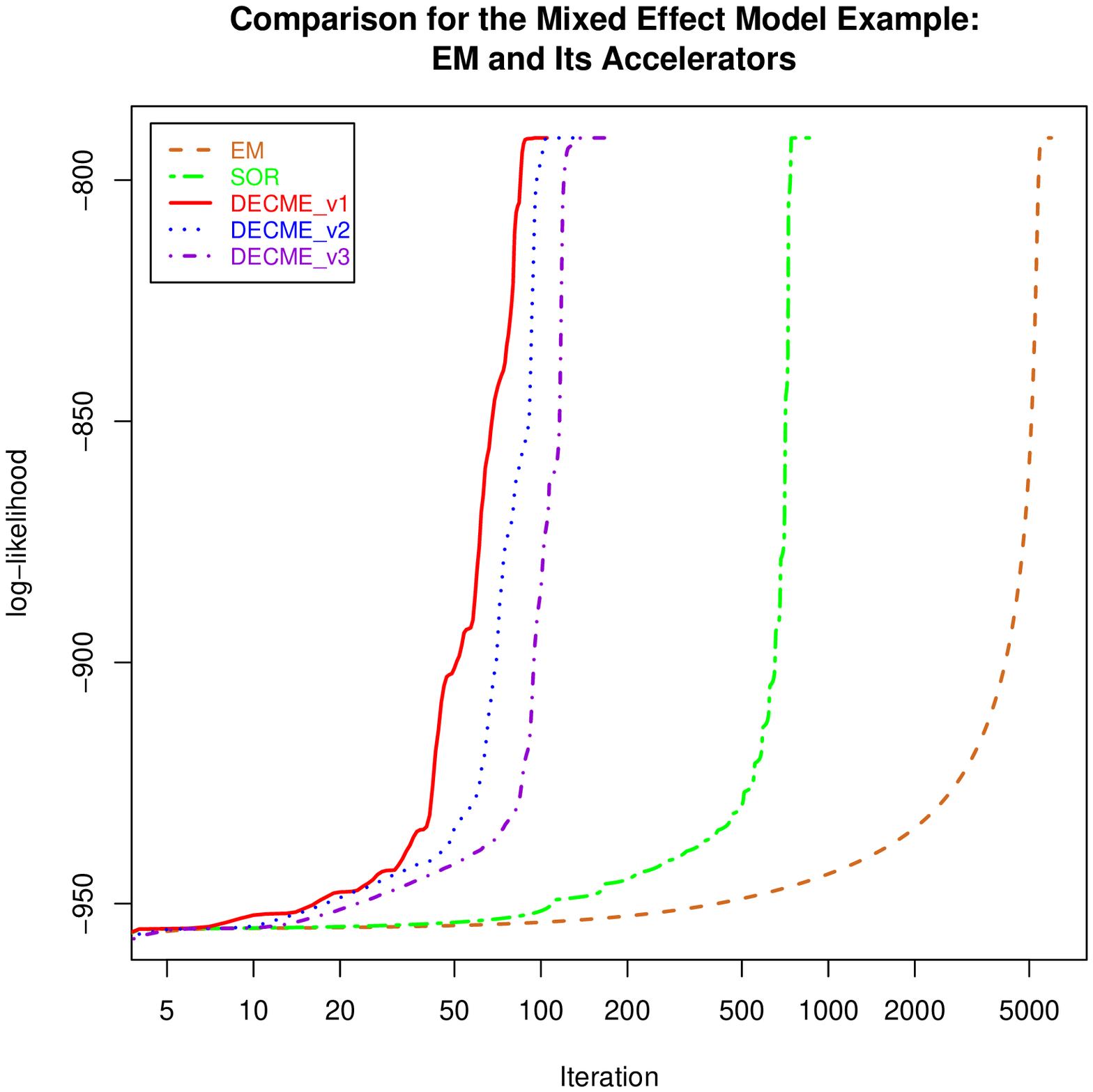}%
 \caption{{\small Comparison for the Linear Mixed-effects Model Example in Section
\ref{sec:rat} and Section \ref{sec:rat_2}. Displayed are increases
in $L(\theta|Y_{obs})$.}} \label{fig:rat_em}
\end{figure}

\begin{figure}
 \centering
\includegraphics[width=3in]{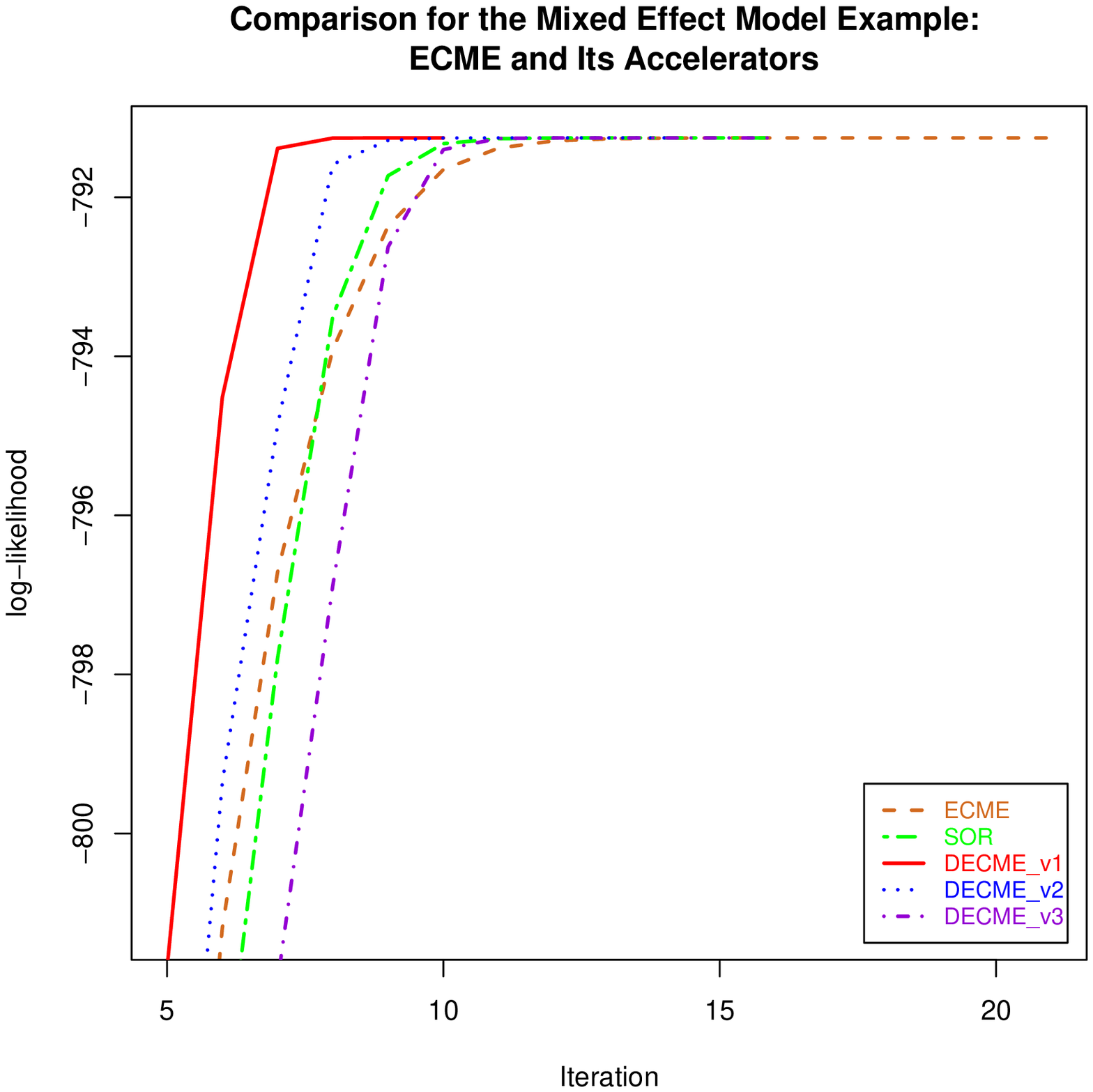}%
 \caption{{\small Comparison for the Linear Mixed-effects Model Example in Section
\ref{sec:rat} and Section \ref{sec:rat_2}. Displayed are increases
in $L(\theta|Y_{obs})$, cont'd}} \label{fig:rat_ecme}
\end{figure}

For this example, ECME converges dramatically faster than EM, as
shown in Figures \ref{fig:rat_em} and \ref{fig:rat_ecme} and Tables
\ref{tbl: iteration} and \ref{tbl: CPU_time}. Specifically, EM takes
$5,968$ iterations and $518.9$ seconds to converge. With the same
setting, ECME (version 1 in \citet{Liu:Rubin:ecme:1994} with
$\theta_{\mathscr{P}_Q}=(\Psi_{11},\Psi_{12},\Psi_{22},\sigma^{2'})'$
and $\theta_{\mathscr{P}_L}=\beta$) uses only 20 iterations and
$1.8$ seconds. The gain of ECME over EM is explained clearly by the
relation between the slow converging directions of EM and the
partition of the parameter space for ECME. From Table \ref{tbl:
rateigenvalue}, the two largest eigenvalues of $DM^{EM}$ are
$0.9860$ and $0.9746$, which are close to 1 and make EM converge
very slow. From Table \ref{tbl: rat}, it is clear that the first
four ``worst" directions of EM fall entirely in the subspace
determined by the fixed effect $\beta$. Since
$\theta_{\mathscr{S}_{L}}=\beta$ for ECME, the slow convergence of
EM induced by the four slowest directions is diminished by
implementing the $ML$-step along the subspace of $\beta$. This is
clear from the row ECME in Table \ref{tbl: rateigenvalue}, where we
see the four largest eigenvalues of $DM^{EM}$ become 0 in
$DM^{ECME}$ while the five small eigenvalues of $DM^{EM}$ remain the
same for $DM^{ECME}$.

\subsection{A Factor Analysis Model Example}\label{sec:battery}
Consider the confirmatory factor analysis model example in
\cite{joreskog:1969}, \cite{Rubin:Thayer:emFA:1982}, and
\cite{Liu:Rubin:fa:1998}. The data is provided in
\citet{Liu:Rubin:fa:1998} and the model is as follows. Let $Y$ be
the observable nine-dimensional variable on an unobservable variable
$Z$ consisting of four factors. For $n$ independent observations of
$Y$, we have
\begin{equation}
Y_i|(Z_i, \beta, \sigma^2) \sim N(Z_i\beta,
\mathsf{diag}(\sigma_1^2,\cdots,\sigma_9^2))
\end{equation}
 where $\beta$ is the $4\times 9$ factor-loading matrix, $\sigma^2=(\sigma_1^2,\cdots,\sigma_9^2)'$
 is called the vector of uniquenesses, and given $(\beta,\sigma^2)$, $Z_1,\cdots, Z_n$ are independently and identically distributed
with $Z_i \sim N(0, I_4), i = 1,\cdots,n$. In the model, there are
zero factor loadings on both factor $4$ for variables 1-4 and on
factor 3 for variables 5-9. Let $\beta_{j\cdot},\ j=1,\cdots,4$ be the four
rows of $\beta$, then the vector of the $36$ free parameters is
$\theta=(\beta_{1\cdot},\beta_{2\cdot},\beta_{3,1-4},\beta_{4,5-9},\sigma^2)'$.
\cite{Liu:Rubin:fa:1998} provided detailed comparison between EM and
ECME. Figure 1 of \cite{Liu:Rubin:fa:1998} shows that
 the gain of ECME over EM is impressive, but not as significant as
ECME for the previous linear mixed-effects model example in Section
\ref{sec:rat}.

The slow convergence of EM for this example is easy to explain from
Table \ref{tbl: batteryeigenvalue} which shows that $DM^{EM}$ has
multiple eigenvalues close to 1. From Table \ref{tbl: battery}, the
eigenvector corresponding to the dominant eigenvalue of $DM^{EM}$
falls entirely in the subspace spanned by $\beta_1$ and $\beta_2$.
This clearly adds difficulty to the ECME version suggested by
\citet{Liu:Rubin:fa:1998} where
$\theta_{\mathscr{S}_{Q}}=(\beta_{1\cdot},\beta_{2\cdot},\beta_{3,1-4},\beta_{4,5-9})'$
and $\theta_{\mathscr{S}_{L}}=\sigma^2$. For this version of ECME, the
eigenvalues of $DM^{ECME}$ are given in row ECME-1 of Table
\ref{tbl: batteryeigenvalue}, where we see that the dominant
eigenvalue of $DM^{EM}$ remains unchanged for $DM^{ECME}$. To
eliminate the effect of the slowest direction of EM, we can try
another version of ECME by letting $\theta_{\mathscr{S}_{Q}}=\sigma^2$
and $\theta_{\mathscr{S}_{L}}=(\beta_{1\cdot}, \beta_{2\cdot}, \beta_{3,1-4},
\beta_{4,5-9})'$. The eigenvalues of $DM^{ECME}$ for this version
are given in row ECME-2 of Table \ref{tbl: batteryeigenvalue}.
Although the second version of ECME is more efficient than the first
version, it is difficult in general to eliminate all the large
eigenvalues in $DM^{EM}$ by accelerating EM in a fixed subspace. For
example, the eigenvector corresponding to the second largest
eigenvalue of $DM^{EM}$ shown in Table \ref{tbl: battery} is not in
the subspace spanned by any subset of the parameters.

\section{The DECME Algorithm}\label{sec:decme_sor}
\subsection{The Generic DECME Algorithm}\label{sec:DECME}
As shown in last section, the efficiency gain of ECME
over its parent EM based on static choices of the acceleration subspaces may be
limited since the slowest converging directions of EM
depend on both the data and model.
It is thus expected to have
a great potential to construct the acceleration subspaces
dynamically based on, for example, the information from past iterations.
This idea is formulated as the following generic DECME algorithm.
At the
$t^{th}$ iteration of DECME, the algorithm proceeds as follows.
\begin{itemize}
\item[] {\bf The Generic DECME Algorithm: the $t^{th}$ iteration}\\
{\bf Input:}  $\tilde{\theta}_{t-1}$\\
 {\large \textbf{E-step:}} Same as the E-step of the original
 EM algorithm;\\
{\large \textbf{M-step:}}
 Run the following two steps:
 \begin{itemize}
 \item[]{\textbf{CM-step:}} Compute $\theta_{t}=\mathsf{argmax}_{\theta}Q(\theta|\tilde{\theta}_{t-1})$
  as in the original EM algorithm;

 \item[]{\textbf{Dynamic CM-step:}} Compute $\tilde{\theta}_{t}= \mathsf{argmax}_{\theta \in
 \mathscr{V}_t}L(\theta|Y_{obs})$,
 where  $\mathscr{V}_t$ is a low-dimensional subspace
 with $\theta_t \in \mathscr{V}_t$.
 \end{itemize}
 \end{itemize}

As noted in \citet{Meng:van:AEM:1997}, the $ML$-steps in ECME should be carried out after the $MQ$-steps to ensure convergence.
Under this condition, ECME with only a single $ML$-step is obviously a special case of DECME. In case multiple $ML$-steps are performed in ECME, a slightly relaxed version of the Dynamic CM-step, {\it i.e.,} simply computing $\tilde{\theta}_{t}$ such that $L(\tilde{\theta}_{t}|Y_{obs})\geq L(\theta_{t}|Y_{obs})$, will still   make DECME a generalisation of ECME.
In either case, the monotone increase of the likelihood function in DECME is
guaranteed by that of the nested EM algorithm \citep{DLR:1977, Wu:1983}, which ensures the stability of DECME. The
convergence rate of DECME relies on the structure of the specific
implementation, {\it i.e.,} how $\mathscr{V}_t$ is constructed.
Furthermore, the well-known method of SOR can be viewed as a special
case of DECME. As shown in Section \ref{sec:sor_decme}, SOR suffers from what is known as the
zigzagging problem and is, thereby, often inefficient. Section
\ref{sec:DECMEv1-3} proposes three efficient alternatives.

\subsection{The SOR method: an Inefficient Special Case of
DECME}\label{sec:sor_decme}
Let $\{\theta_{t}-\tilde{\theta}_{t-1}\}$ represent the linear
subspace spanned by $\theta_{t}-\tilde{\theta}_{t-1}$. SOR can be
obtained by specifying $\mathscr{V}_t = \theta_{t}+
\{\theta_{t}-\tilde{\theta}_{t-1}\}$ in the Dynamic CM-step of
DECME, {\it i.e.}, $\tilde{\theta}_{t}=\theta_{t}+\alpha_{t}d_{t}$,
$d_{t}=\theta_{t}-\tilde{\theta}_{t-1}$, and
$\alpha_t=\mathsf{argmax}_{\alpha}L(\theta_{t}+ \alpha
d_{t}|Y_{obs})$. The so-called relaxation factor $\alpha_t$ can be
obtained by a line search. See Figure \ref{fig:illustration_decme}
for an illustration of the SOR iteration.

The reason that SOR may be used to accelerate EM is clear from the
following theorem which implies that, in a small neighbourhood of the
MLE, a point with larger likelihood value can always be found by
enlarging the step size of EM:
\begin{thm}
\label{thm:sor_ss} In a small neighbourhood of $\hat{\theta}$,
 the relaxation factor $\alpha_t$ of SOR is always positive.
\end{thm}

The proof is given in Appendix \ref{app:EM} and the conservative
movement of EM is illustrated in Figure \ref{fig:path} for a
two-dimensional simulated example. For simplicity, it has also been
proposed to choose $\alpha_t$ as a fixed positive number \citep[{\it
e.g.,} ][]{Lang:1995b}. We call this version with fixed $\alpha_t$
the SORF method. Let $\lambda_1$ and $\lambda_p$ be the largest and
smallest eigenvalues of $I_{com}^{-1}I_{obs}$ (see Appendix
\ref{app:EM} for detailed discussion). It is well known that SORF achieves its
optimal convergence rate
$(\lambda_1-\lambda_p)/(\lambda_1+\lambda_p)$ if
$\alpha_t=2/(\lambda_1+\lambda_p)-1$ for any $t$ \citep[see, {\it
e.g.},][]{Salakhutdinov03adaptiveoverrelaxed}. In the past, the
theoretical argument for SOR has been mainly based on this fact,
which is obviously insufficient. The following theorem provides new
angles for understanding the convergence of SOR.
\begin{thm}
\label{thm:sor} For a two-dimensional problem ({\it i.e.,} $p=2$) and in a small neighbourhood of
$\hat{\theta}$, the following results hold for SOR:
 \newcounter{thmsor}
\begin{list}
{\arabic{thmsor}.)}{\usecounter{thmsor}\itemsep=0cm}
\item $\alpha_{t}= \alpha_{t-2}$;
%\item $[1-(1+\alpha_{t})\lambda_1][1-(1+\alpha_{t-1})\lambda_1]=[1-(1+\alpha_{t})\lambda_2][1-(1+\alpha_{t-1})\lambda_2]$
\item SOR converges at least as fast as the optimal SORF, and the optimal SORF
 converges faster than EM; and
\item SOR oscillates around the slowest converging direction of EM; The SOR estimates from the odd-numbered iterations lie on the same line and so do those from the even-numbered iterations; Furthermore, the two lines intersect at the MLE $\hat{\theta}$.
\end{list}
\end{thm}

The proof is provided in Appendix \ref{app:SOR:convergence}. The zigzagging phenomena of SOR revealed by conclusion 3 is illustrated in Figure \ref{fig:path}. For the case of
$p>2$, it is interesting to see that the relaxation factors
$\alpha_t$ generated from SOR also have a similar oscillating
pattern as that for $p=2$ (conclusion 1). This is illustrated in
Figure \ref{fig:relaxationfactor}. The top panel of Figure
\ref{fig:relaxationfactor} shows the relaxation factors for the
two-dimensional example used to generate Figure \ref{fig:path} and
the lower panel shows those for a nine-dimensional simulated
example. The nine-dimensional example is generated by simulating the
behaviour of EM in a small neighbourhood of the MLE for the linear mixed-effects model example in Section \ref{sec:rat} and Section
\ref{sec:rat_2}.

\subsection{DECME\_v1 and its Variants: Three Efficient DECME Implementations}\label{sec:DECMEv1-3}
\subsubsection{The Basic Version: DECME\_v1}
\begin{figure}
 \centering
\includegraphics[width=3in]{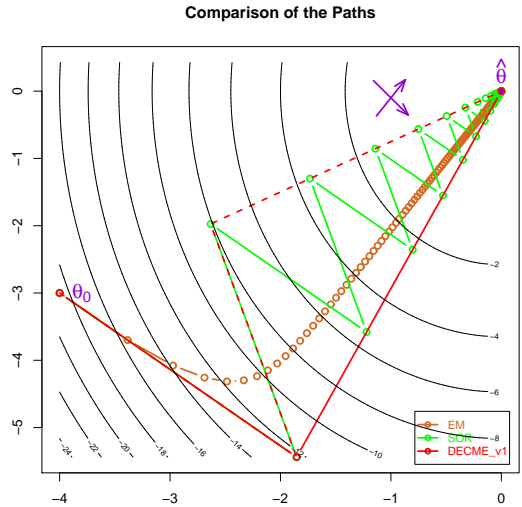}%
\caption{{\small Comparison of the Paths of EM, SOR, and DECME\_v1
for a Two-dimensional Example. The eigenvalues of $DM^{EM}$ are
$0.9684$ and $0.6232$; the darkviolet cross on the upright corner
shows the directions of the two eigenvectors of $DM^{EM}$; The red
dashed lines represent the true path of DECME\_v1 in its second
iteration.}} \label{fig:path}
\end{figure}

\begin{figure}
 \centering
\includegraphics[width=3in]{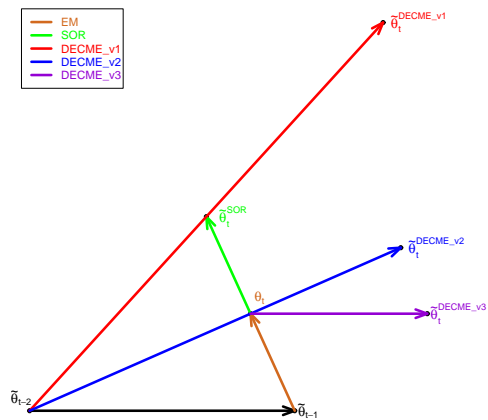}%
\caption{{\small Illustration for One Iteration of the DECME
Implementations.}} \label{fig:illustration_decme}
\end{figure}

\begin{figure}
 \centering
\includegraphics[width=3in]{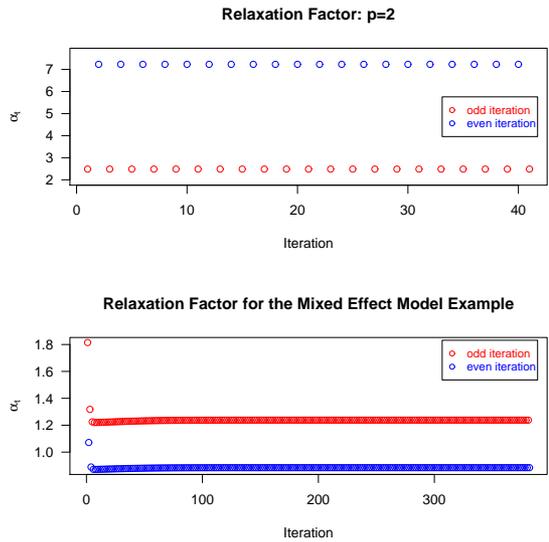}%
\caption{{\small Relaxation Factor $\alpha_t$ Generated from SOR.
The top panel plots the sequence of $\alpha_t$ for the
two-dimensional example used to generate Figure \ref{fig:path}, and
the bottom panel plots the sequence of $\alpha_t$ from the simulated
nine-dimensional example in Section \ref{sec:sor_decme} by using
information from the linear mixed-effects model example in Section
\ref{sec:rat} and Section \ref{sec:rat_2} .}}
\label{fig:relaxationfactor}
\end{figure}

The zigzagging problem has long been considered to be one of the major
disadvantages for optimisation algorithms since the effective
movement towards the MLE is usually small even if the step size is
large.
% Conclusion 3 of Theorem \ref{thm:sor} suggests a line search
Figure \ref{fig:path} suggests a line search along the line
connecting the zigzag points, as shown by one of the red dashed
lines. For two-dimensional quadratic functions, this suggested
procedure shown in Figure \ref{fig:path} converges immediately.
Although this only represents a very rare case in practise, it
motivated us to consider efficient DECME implementations.

One way to proceed is to repeat the procedure shown by the red
dashed lines in Figure \ref{fig:path}, {\it i.e.,} each cycle of the
new algorithm includes two iterations of SOR and a line search along
the line connecting the initial point of the current cycle and the
end point of the second SOR iteration. Numerical experiments show
that this procedure is not very effective.

Another way to proceed is what we call DECME\_v1. DECME\_v1 retains
the procedure shown by the red dashed lines in Figure \ref{fig:path}
as its first two iterations and is formally defined as follows. At
the first iteration of DECME\_v1, $\tilde{\theta}_1$ is obtained by
running one iteration of SOR from the starting point
$\tilde{\theta}_0$. At the $t^{th}$ iteration of DECME\_v1, one
iteration of SOR is first conducted to obtain
$\tilde{\theta}^{SOR}_{t}$, followed by a line search along the
line connecting $\tilde{\theta}_{t-2}$ and
$\tilde{\theta}^{SOR}_{t}$ to obtain $\tilde{\theta}_{t}$. The
process is continued for $p$ iterations and restarted with a
standard SOR iteration. The reason for restarting becomes clear from
Theorem \ref{thm:AEM:DECME} below, which shows that DECME\_v1 is
equivalent to a conjugate direction method.

Formally, DECME\_v1 is described in the framework of the generic DECME
algorithm by implementing the Dynamic CM-step with two line searches as
follows (except for the iterations where the algorithm is
restarted):
\begin{itemize}
\item[] {\bf Dynamic CM-step of DECME\_v1: the $t^{th}$ iteration}
\begin{itemize}
 \item[] \textbf{Substep 1:} Calculate
 $\tilde{\theta}^{SOR}_{t}=\theta_{t}+\alpha_{t}^{(1)}d_t^{(1)}$,
 where $d_t^{(1)}=\theta_{t}-\tilde{\theta}_{t-1}$, and $\alpha_{t}^{(1)}=\mathsf{argmax}_{\alpha}L(\theta_{t}+\alpha d_t^{(1)}|Y_{obs})$;
 \item[] \textbf{Substep 2:} Calculate $\tilde{\theta}_{t}=\tilde{\theta}^{SOR}_{t} + \
 \alpha_{t}^{(2)}d_t^{(2)}$, where $d_t^{(2)}=\tilde{\theta}^{SOR}_{t}-\tilde{\theta}_{t-2}$, and
$\alpha_{t}^{(2)}=\mathsf{argmax}_{\alpha}L(\tilde{\theta}^{SOR}_{t}+\alpha d_t^{(2)}|Y_{obs})$.
 \end{itemize}
 \end{itemize}

\noindent An illustration of the DECME\_v1 iteration is given in Figure \ref{fig:illustration_decme}. We note that $\tilde{\theta}_{t}$ is actually the point
that maximises $L(\theta|Y_{obs})$ over the two-dimensional subspace
$\mathscr{V}_t=\tilde{\theta}_{t-1}+\{
\tilde{\theta}_{t-1}-\tilde{\theta}_{t-2}, \
\theta_{t}-\tilde{\theta}_{t-1}\}$ under certain conditions. This
can be seen from the proof, given in Appendix \ref{app:AEM:DECME},
of the
following theorem,% \ref{thm:AEM:DECME}
which demonstrates the efficiency of DECME\_v1.

\begin{thm}
\label{thm:AEM:DECME} In a small neighbourhood of the MLE, DECME\_v1
with exact line search is equivalent to the conjugate direction
method AEM.
\end{thm}

Theorem \ref{thm:AEM:DECME} implies that DECME\_v1 is about as
efficient as AEM  near the MLE in terms of the number of iterations.
As noted in Section 1, DECME\_v1 is much easier to implement and can be
made automatic for almost all EM algorithms, which typically have
coded likelihood evaluation routines for debugging code and
monitoring convergence. A line search method is needed for DECME\_v1,
but can be implemented once for all; whereas evaluation of the
gradient vector of $L(\theta|Y_{\rm obs})$ required for AEM is
problem specific and thus demands substantially more programming
efforts. Note also that gradient evaluation can be expensive.
For example, for comparing different methods in the optimisation
literature, it is often to count one evaluation of the gradient
vector as $p$ function evaluations, where $p$ stands for the
dimensionality of the parameter space.

The idea behind DECME\_v1 is very similar to the parallel tangent
(PARTAN) method for accelerating the steepest descent method
\citep{Shah:Buehler:Kempthorne:1964}. PARTAN can be viewed as a
particular implementation of the conjugate gradient method
\citep{Fletcher:Reeves:1964}, developed based on the method in
\citet{Hestenes:Stiefel:1952} for solving linear systems. It is also
worth noting that PARTAN has certain advantage over the conjugate
gradient method as discussed in \citet[p.~257]{luenberger:2003}. For
example, the convergence of PARTAN is more reliable than the
conjugate gradient method when inexact line search is used as is
often the case in practise.

\subsubsection{DECME\_v2 and DECME\_v3} \label{sec:decme_v23}
DECME\_v1 requires two line search steps in one iteration. DECME\_v2 and DECME\_v3, the two variants of DECME\_v1 that require a single line search in each iteration, are obtained by specifying a one-dimensional acceleration subspace in the dynamic CM-step as  $\mathscr{V}_{t}=\theta_{t}+\{\theta_{t}-\tilde{\theta}_{t-2}\}$ and
$\mathscr{V}_{t}=\theta_{t}+\{\tilde{\theta}_{t-1}-\tilde{\theta}_{t-2}\}$, respectively.
This is depicted in Figure \ref{fig:illustration_decme}.

There is no much difference among the three newly proposed methods
and SOR in terms of programming since their main building blocks,
the EM iteration and a line search scheme, are the same. However, SOR
can hardly compete with the new methods for all the examples we have
observed. Among the three new implementations, DECME\_v1 usually uses the smallest number of iterations to
converge while it obviously takes more time
to run one DECME\_v1 iteration. Hence when the cost of running a line search, determined
mainly by the cost of computing the log-likelihood, is low relative
to the cost of running one EM iteration, DECME\_v2 and DECME\_v3 may
be more efficient than DECME\_v1 in terms of CPU time. These points are shown by the examples in next
section.

\section{Numerical Examples}\label{sec:numerical}

In this section we use four sets of numerical examples to compare
the convergence speed of EM, SOR, DECME\_v1, DECME\_v2, and
DECME\_v3 in terms of both number of iterations and CPU time.

\subsection{The Setting for the Numerical Experiments} \label{sec:setting}
The line search scheme for all the examples is implemented by making
use of the {\it optimize} function in R. The detailed discussion
about the configuration of the function and how to achieve line
search for constrained problems with line search routines designed
for unconstrained problems is given in Appendix
\ref{app:line_search}.

For the three examples in Section \ref{sec:rat_2} (Section
\ref{sec:rat}), \ref{sec:battery_2} (Section \ref{sec:battery}), and \ref{sec:bt}, we first run
EM with very stringent stopping criterions to obtain the maximum
log-likelihood $l_{max}$ for each example. Then we run each of the
five (nine for the example in Section \ref{sec:rat_2}) different
algorithms from the same starting point and terminate them when
$L(\theta|Y_{obs})$ is not less than $l_{max}-10^{-6}$. The results
for these three examples, including number of iterations and CPU
time, are summarised in Table \ref{tbl: iteration} and \ref{tbl:
CPU_time}. The increases in $L(\theta|Y_{obs})$ against the number
of iteration for each of the three examples are shown in Figures
\ref{fig:rat_em}, \ref{fig:rat_ecme}, \ref{fig:battery} and
\ref{fig:bt}. The setting and the results for the simulation study
in Section \ref{sec:gm} are slightly different.

\subsection{The Linear Mixed-effects Model Example}
\label{sec:rat_2}

The mixed-effects model example of Section \ref{sec:rat} is used to
illustrate the performance of DECME when applied to accelerate both
EM and ECME. The increases of $L(\theta|Y_{obs})$ are shown in
Figure \ref{fig:rat_em} and \ref{fig:rat_ecme}. For this example,
while EM needs $5,968$ iterations and SOR needs $918$ iterations to
converge, all three new implementations of DECME needs no more than 170
iterations with only 104 iterations for DECME\_v1. In terms of CPU
time all three new methods converge about 30 times faster
than EM. It is also interesting to see that the new methods works very well
for accelerating ECME (especially DECME\_v1 further reduces the
number of iterations from 20 to 9) even when ECME already converges
much faster than EM.

\subsection{The Factor Analysis Model Example} \label{sec:battery_2}
\begin{figure}
 \centering
\includegraphics[width=3in]{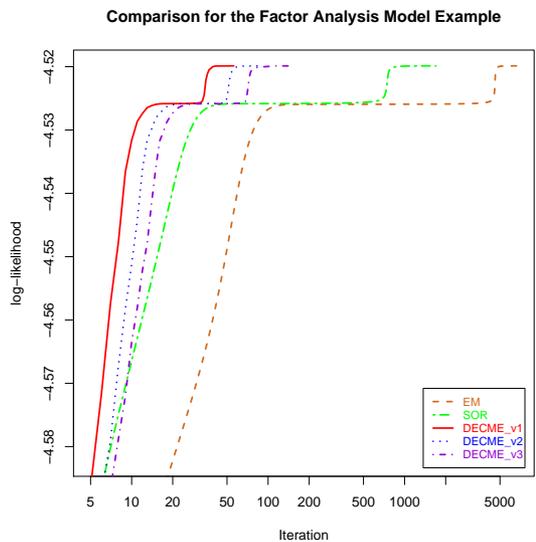}%
\caption{{\small Comparison for the Factor Analysis Model Example in
Section \ref{sec:battery} and \ref{sec:battery_2}. Displayed are
increases in $L(\theta|Y_{obs})$.}} \label{fig:battery}
\end{figure}

\begin{figure}
\centering
\includegraphics[width=3in]{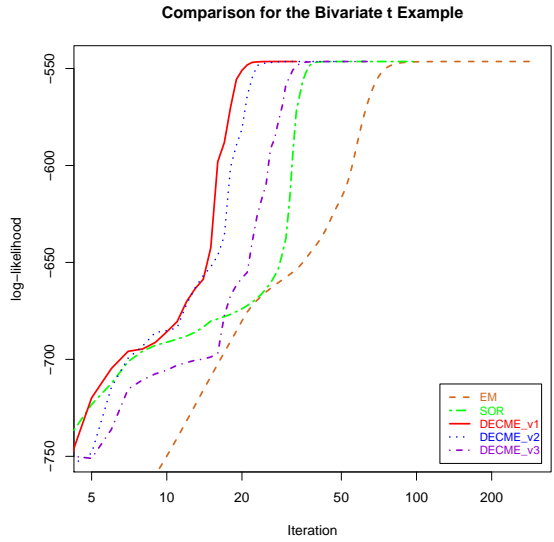}
 \caption{{\small Comparison for the
Bivariate $t$ Example in Section \ref{sec:bt}. Displayed are
increases in $L(\theta|Y_{obs})$.}} \label{fig:bt}
\end{figure}

The factor analysis example of Section \ref{sec:battery} and the
same starting point used in \citet{Liu:Rubin:fa:1998} are used here.
The increases of $L(\theta|Y_{obs})$ are shown in Figure
\ref{fig:battery}. For this example, while all three new implementations of
DECME converges much faster than EM and SOR, DECME\_v1 uses only
less than 1\% of the number of iterations of EM (55 to 6,672) and is
about 30 times faster than EM in terms of CPU time (1.2 seconds {\it
vs.} 33.4 seconds). It is also interesting to see that all three new
methods pass the flat period shown in Figure
\ref{fig:battery} much more quickly than both EM and SOR. As
discussed in \citet{Liu:Rubin:fa:1998}, the long flat period of EM
and SOR before convergence makes it difficult to assess the
convergence. Clearly, this does not appear to be a problem for the
three new methods.

\subsection{A Bivariate $t$ Example} \label{sec:bt}
Let $t_{p}(\mu,\Psi,\nu)$ represent a multivariate $t$ distribution
with $\mu$, $\Psi$, and $\nu$ as the mean, the covariance matrix,
and the degree of freedom, respectively. Finding the MLE of the
parameters $(\mu,\Psi,\nu)$ is a well known interesting application
of the EM-type algorithms.

Here we use the bivariate $t$ distribution example in
\cite{Liu:Rubin:ecme:1994}, where the data is adapted from Table 1
of \cite{Cohen:Dalal:Tukey:robu:1993}. Figure \ref{fig:bt} shows the
increases of $L(\theta|Y_{obs})$ for each algorithm starting from
the same point $(\mu,\Psi,\nu)=((0,0)', \mathsf{diag}(1,1),1)$. For
this example, EM converges relatively fast with 293 iterations and
1.4 seconds. But we still see the advantage of the new implementations of
DECME over EM and SOR: DECME\_v1 uses only 32 iterations and 0.9 second
to converge. Also note that SOR uses 1.5 seconds to converge,
which is slightly more than that of EM.

\subsection{Gaussian Mixture Examples}\label{sec:gm}
The EM algorithm is wildly acknowledged as a powerful method for
fitting the mixture models, which are popular in many different
areas such as machine learning and pattern recognition \citep[{\it
e.g.,} ][]{Jordan:Robert:1994, McLachlan:Krishnan:1997,
McLachlan:Peel:2000, Bishop:2006}. While the slow convergence of EM
has been frequently reported for fitting mixture models, a few
extensions have been proposed for specifically accelerating this EM
application \citep[among others]{Liu:Sun:ecme:mixture:1997,
Dasgupta:Schulman:2000, Celeux:Chretien:Forbes:2001,
Pilla:Lindsay:2001}. Here we show that DECME, as an off-the-shelf
accelerator, can be easily applied to achieve dramatically faster
convergence than EM.

\begin{figure}
\centering
\includegraphics[width=3in]{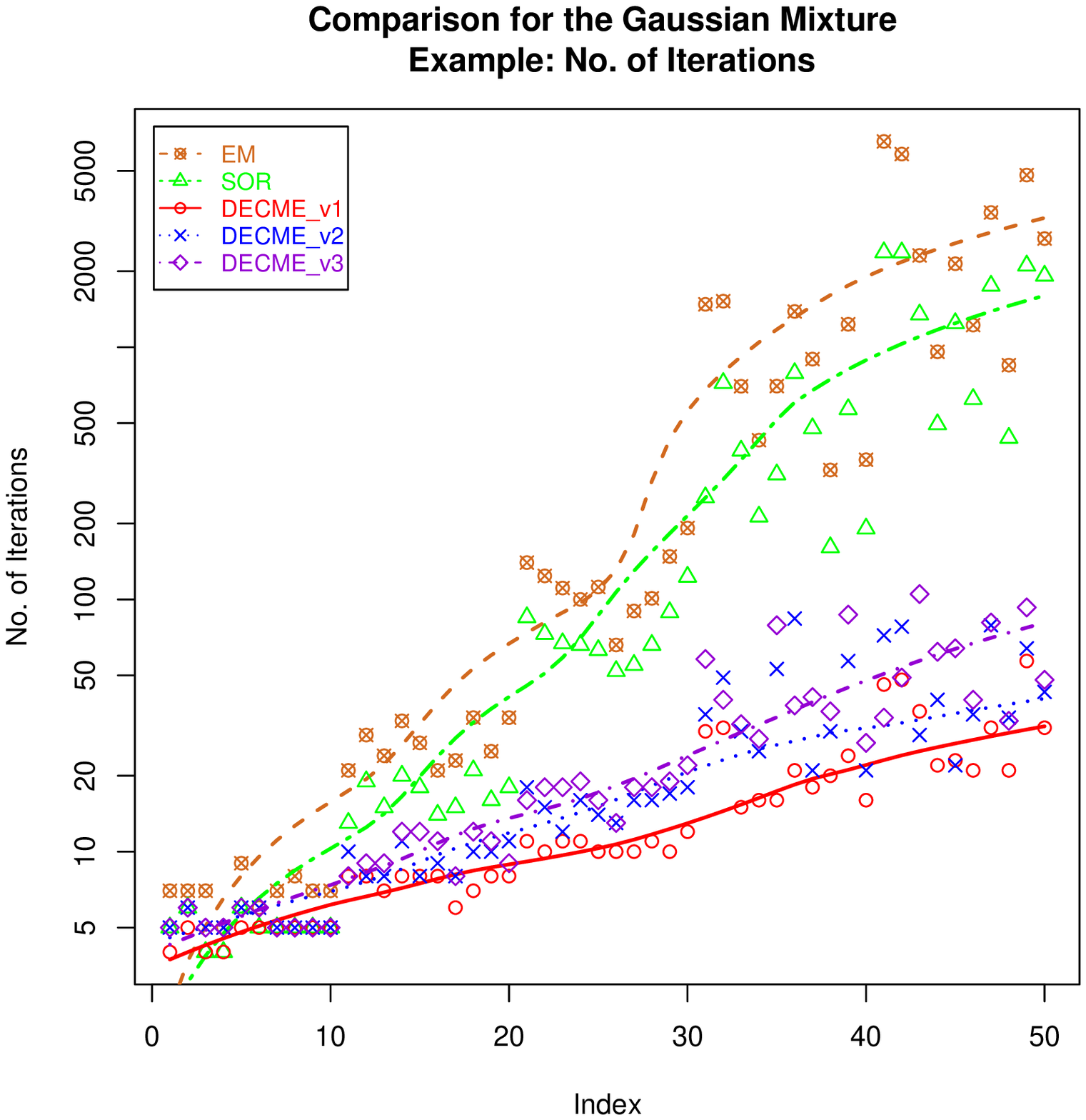}%
 \caption{{\small Comparison for the
Gaussian Mixture Example in Section \ref{sec:gm}. Displayed are
number of iterations; the smoothed curves are generated by robust
local regression \citep{Cleveland:1979}.}} \label{fig:gm_1}
\end{figure}

\begin{figure}
\centering
\includegraphics[width=3in]{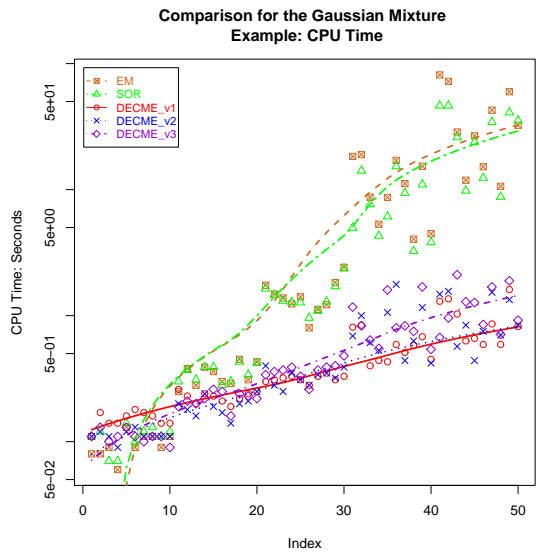}%
 \caption{{\small Comparison for the
Gaussian Mixture Example in Section \ref{sec:gm}. Displayed are CPU
time; the smoothed curves are generated by robust local
regression.}} \label{fig:gm_2}
\end{figure}

A class of mixtures of two univariate normal densities is used to
illustrate the relation between the efficiency of EM and the
separation of the component populations in the mixture in
\cite{Redner:Walker:1984}. Specifically, the mixture has the form of
\begin{equation}
\begin{array}{rcl}
p(x|\pi_1, \pi_2, \mu_1, \mu_2, \sigma_1^2, \sigma_2^2)&=&\pi_1
p_1(x|\mu_1, \sigma_1^2)+\pi_2
p_2(x|\mu_2, \sigma_2^2),\\
p_i(x|\mu_i, \sigma_i^2)&=&\frac{1}{\sqrt{2\pi
\sigma_i}}e^{-(x-\mu_i)^2/2\sigma_i^2}, \ i=1,2.
\end{array}
\end{equation}
Let $\pi_1=0.3, \ \pi_2=0.7, \ \sigma_1^2=\sigma_2^2=1$, and
$\mu_1=-\mu_2$, then ten random samples of $1,000$ observations were
 generated from each case of $\mu_1-\mu_2=6,\ 4,\ 3,\ 2$ and $1.5$.
We ran EM, SOR, DECME\_v1, DECME\_v2, and DECME\_v3 from the same
starting point $\pi_1^{(0)}=\pi_2^{(0)}=0.5, \
\sigma_i^{2^{(0)}}=0.5,\ \mu_i^{(0)}=1.5\mu_i$. The algorithms are
terminated when $||\theta_{t+1}-\theta_{t}||_1\ < \ 10^{-5}$, where
$||\cdot||_1$ represents the $l_1$ norm. The results for number of
iterations and CPU time are shown in Figure \ref{fig:gm_1} and
Figure \ref{fig:gm_2}, respectively.

We see that SOR typically uses about half of the number of
iterations of EM while the two have very similar performance in
terms of CPU time. When EM converges very slowly, the three new
implementations of DECME can be dramatically faster than EM with a factor
100 or more in terms of number of iterations and a factor of 50 or
more in terms of CPU time. When EM converges very fast, from Figure
\ref{fig:gm_2}, we see some overlapping among several methods in
terms of CPU time for the first group of ten simulations, although
all the accelerators still outperform EM in terms of number of
iterations. In practise, fast convergence like this is somewhat rare
for EM. Notice that all the methods take less than one second to
converge. Hence, this phenomenon of overlapping should not be used to
dismiss the advantage of the accelerators. Nevertheless, this may
serve as empirical evidence to support the idea of accelerating EM
only after a few EM iterations have been conducted as suggested in
\citet{Jamshidian:1993}.

\section{Discussion}\label{sec:discussion}
The limitation of using fixed acceleration subspaces in ECME led to
the idea of dynamically constructing the subspaces for the
supplementary $ML$-steps. We formulated this idea as the generic
DECME algorithm, which provides a simple framework for developing stable and efficient acceleration methods of EM. The zigzagging problem of SOR, a special case of
DECME, motivated the development of the three new DECME
implementations, {\it i.e.,} DECME\_v1-v3. The stability of DECME is
guaranteed by the nested EM iteration. The equivalence of DECME\_v1
to AEM, a conjugate direction method, provides theoretical
justification for the fast convergence of the new methods, which is also supported by the numerical results. Moreover, the simplicity of the new methods makes them more attractive than AEM.

In optimisation literature, it is popular to analyse the convergence of optimisation algorithms near the optimum with the assumption of an ideal exact line search. However, exact line search is not a realistic choice in practise (see the discussion in
Appendix \ref{app:line_search}). Hence, the relative performance of DECME\_v1 and AEM,
in both efficiency and stability, could be very different, especially when the starting point is far from the MLE. Note also that many works have been done to expedite the line search for SOR, {\it e.g.,} \cite{Salakhutdinov03adaptiveoverrelaxed} and
\cite{Hesterberg:stag:2005}. It will be interesting to see how the
similar techniques perform for DECME for we have shown that the
newly proposed acceleration directions work much better than that
used by SOR.

Our main focus in the current paper has been on accelerating EM.
However, it is noteworthy that the proof of Theorem \ref{thm:AEM:DECME}
 only depends on the linear convergence rate of the underlying algorithm
 being accelerated rather than its specific structure. Hence an immediate
 point to make is that the new methods should also work for other EM-type
 algorithms of linear convergence rate or more broadly for the MM algorithm \citep{hunter:lange:2004}. We leave this problem open for future investigation.

\section*{Acknowledgement}
The authors thank Mary Ellen Bock, Aiyou Chen, William Cleveland,
Ahmed Sameh, David van Dyk, Hao Zhang, Heping Zhang, and Jian Zhang
for helpful discussions and suggestions. Yunxiao He's research was
partially supported by NIH grant R01DA016750.

\section*{Appendix}
\appendix
\section{The Linear Convergence Rate of EM: a Quick Review} \label{app:review}
This section reviews some well known convergence properties of EM to
establish necessary notations. These results are mainly adapted from
\citet{DLR:1977} and \citet{Meng:Rubin:convergenceEM:1994}.

 In a small neighbourhood of the MLE, the
observed log-likelihood $L(\theta | Y_{obs})$ may be assumed to be a
quadratic function:
\begin{equation}
\label{equ:obslik}
L(\theta|Y_{obs})=-\frac{1}{2}(\theta-\hat{\theta})'I_{obs}(\theta-\hat{\theta}).
\end{equation}
Under this assumption, \citet{DLR:1977} proved that EM has a linear
convergence rate determined by $DM^{EM}$, {\it i.e.,} equation (\ref{equ:EM}). We
mentioned previously that $DM^{EM}$ is called the missing
information fraction. It is named after the following identity:
\begin{equation}
\label{equ:DM}
 DM^{EM}=I_p-I_{com}^{-1}I_{obs}=I_{com}^{-1}I_{mis},
\end{equation}
where $I_p$ represents the identity matrix of order $p$, $I_{obs}$ and $I_{com}$ are the negative Hessian matrices of
$L(\theta|Y_{obs})$ and $Q(\theta|Y_{obs},\hat{\theta})$ at the MLE, and $I_{mis}= I_{com}-I_{obs}$.
The matrices $I_{obs}$, $I_{mis}$ and $I_{com}$ are usually called
observed-data, missing-data, and complete-data information matrices.
We assume that these matrices are positive definite in this
paper.

Since $I_{com}$ is positive definite, there exists a positive
definite matrix, denoted by $I_{com}^{1/2}$, such that
$I_{com}=I_{com}^{1/2}I_{com}^{1/2}$. Further denote by
$I_{com}^{-1/2}$ the inverse of $I_{com}^{1/2}$. Then
$I_{com}^{-1}I_{obs}$ is similar to $I_{com}^{1/2}\times
I_{com}^{-1}I_{obs} \times I_{com}^{-1/2}=
I_{com}^{-1/2}I_{obs}I_{com}^{-1/2}$ and, thereby,
$I_{com}^{-1}I_{obs}$ and $I_{com}^{-1/2}I_{obs}I_{com}^{-1/2}$ have
the same eigenvalues. Since $I_{com}^{-1/2}I_{obs}I_{com}^{-1/2}$ is
symmetric, there exists
 an orthogonal matrix $T$ such that
 \begin{equation}
 \label{equ:principle axis}
I_{com}^{-1/2}I_{obs}I_{com}^{-1/2}= T \Lambda T^{'},
\end{equation}
 where $\Lambda=\mathsf{diag}(\lambda_1, \cdots,
\lambda_p)$, and $\lambda_i,\ i=1,\cdots,p$, are the eigenvalues of
$I_{com}^{-1}I_{obs}$. Therefore,
\begin{equation}
\label{equ:ICIOdecom}
 I_{com}^{-1}I_{obs}=I_{com}^{-1/2}T \Lambda T^{'}I_{com}^{1/2}.
\end{equation}
 Let $P=I_{com}^{-1/2}T$, then we have $I_{com}^{-1}I_{obs}=P \Lambda P^{-1}$. Furthermore, the columns of $P$
and the rows of $P^{-1}$ are eigenvectors of $I_{com}^{-1}I_{obs}$
and $I_{obs}I_{com}^{-1}$, respectively. Define
$\eta =P^{-1}(\hat{\theta}-\theta)$, then from equation (\ref{equ:EM}) we
have $\eta_{t}=(I_p-\Lambda) \eta_{t-1}$, or equivalently
\begin{equation}
\label{equ:eigendir} \eta_{t,i}=(1-\lambda_i) \eta_{t-1,i}, \
i=1,\cdots,p.
\end{equation}
Equation (\ref{equ:eigendir}) implies that EM converges
independently along the $p$ eigenvector directions of
$I_{com}^{-1}I_{obs}$ (or equivalently $DM^{EM}$) with the rates
determined by the corresponding eigenvalues. For simplicity of the
later discussion, we assume $1>\lambda_1
> \lambda_2 > \cdots > \lambda_p > 0$ and $\eta_{0,i}\neq 0, \ i=1,\cdots,p$.

\section{The Conservative Step Size of EM: Proof of Theorem \ref{thm:sor_ss}} \label{app:EM}
From equation (\ref{equ:obslik}) and the definition of SOR in
Section \ref{sec:sor_decme}, it is easy to show that
\begin{equation}
%\label{equ:alpha}
\alpha_{t}=\frac{(\theta_{t}- \tilde{\theta}_{t-1})'I_{obs}({\hat{\theta}}-\theta_{t})}{(\theta_{t}- \tilde{\theta}_{t-1})'I_{obs}(\theta_{t}- \tilde{\theta}_{t-1})}.
\end{equation}
Then making use of the fact that $ \theta_{t}= \theta_{t-1}+
I_{com}^{-1}I_{obs}(\hat{\theta}-\theta_{t-1})$ (followed from
equations \ref{equ:EM} and \ref{equ:DM}) leads to
\begin{equation}
\alpha_{t}
=\frac{({\hat{\theta}}-\tilde{\theta}_{t-1})'I_{obs}I_{com}^{-1}I_{obs}({\hat{\theta}}-\tilde{\theta}_{t-1})}
    {({\hat{\theta}}-\tilde{\theta}_{t-1})'I_{obs}I_{com}^{-1}I_{obs}I_{com}^{-1}I_{obs}({\hat{\theta}}-\tilde{\theta}_{t-1})}-1.
\end{equation}
By definition of $\eta$, we have
${\hat{\theta}}-\tilde{\theta}_{t-1}=I_{com}^{-1/2}T\tilde{\eta}_{t-1}$.
Making use of equation (\ref{equ:principle axis}) and the fact that
$T$ is an orthogonal matrix yields
\begin{equation}
\label{equ:alpha_sor_1} \alpha_{t}
=\frac{\tilde{\eta}_{t-1}'
\Lambda^2 \tilde{\eta}_{t-1}} {\tilde{\eta}_{t-1}' \Lambda^3
\tilde{\eta}_{t-1}}-1.
\end{equation}
Since $\Lambda$ is diagonal and all its diagonal elements are
between $0$ and $1$, it follows immediately that $\alpha_t > 0$.
{\hfill} $\Box$

\section{The Convergence of SOR: Proof of Theorem \ref{thm:sor}}
\label{app:SOR:convergence} Similar to equation (\ref{equ:EM}) and
(\ref{equ:eigendir}) for EM, we have the following results for SOR:
 \begin{equation}
\label{equ:sor} \hat{\theta}-\tilde{\theta}_{t}= [I_p-(1+\alpha_t)
I_{com}^{-1}I_{obs}](\hat{\theta}-\tilde{\theta}_{t-1}),
 \end{equation}
 and
 \begin{equation}
\label{equ:eta}
\tilde{\eta}_{t,i}=[1-(1+\alpha_{t})\lambda_i]\tilde{\eta}_{t-1,i}\
, \ \ i=1,\cdots,p.
\end{equation}

For $p=2$, from equation (\ref{equ:alpha_sor_1}), we have
\begin{equation}
\label{equ:alpha_sor_2} \alpha_{t}=\frac{\lambda_1^2
\tilde{\eta}_{t-1,1}^2+\lambda_2^2
\tilde{\eta}_{t-1,2}^2}{\lambda_1^3
\tilde{\eta}_{t-1,1}^{2}+\lambda_2^3 \tilde{\eta}_{t-1,2}^2}-1,
\end{equation}
and then,
\begin{equation}
\label{equ:rate} 1-(1+\alpha_{t})\lambda_1=\frac{\lambda_2^2
(\lambda_2-\lambda_1)\tilde{\eta}_{t-1,2}^2}{\lambda_1^3
\tilde{\eta}_{t-1,1}^{2}+\lambda_2^3 \tilde{\eta}_{t-1,2}^2},
\
1-(1+\alpha_{t})\lambda_2=\frac{\lambda_1^2
(\lambda_1-\lambda_2)\tilde{\eta}_{t-1,1}^2}{\lambda_1^3
\tilde{\eta}_{t-1,1}^{2}+\lambda_2^3 \tilde{\eta}_{t-1,2}^2}.
\end{equation}
From equation (\ref{equ:eta}) and equation (\ref{equ:rate}), we have
\begin{equation}
\label{equ:eta_2} \frac{\tilde{\eta}_{t,1}}{\tilde{\eta}_{t,2}}
=-\frac{\lambda_2^2}{\lambda_1^2}\frac{\tilde{\eta}_{t-1,2}}{\tilde{\eta}_{t-1,1}}.
\end{equation}
It follows that
$\tilde{\eta}_{t,1}/\tilde{\eta}_{t,2}=\eta_{t-2,1}/\tilde{\eta}_{t-2,2}.$
Furthermore, from equation (\ref{equ:alpha_sor_2}), we have
\begin{equation}
\label{equ:alpha_sor_3}
\alpha_{t}
=\frac{\lambda_1^2(\tilde{\eta}_{t-1,1}/\tilde{\eta}_{t-1,2})^2+\lambda_2^2}{\lambda_1^3
(\tilde{\eta}_{t-1,1}/\tilde{\eta}_{t-1,2})^2+\lambda_2^3}-1,
\end{equation}
and immediately $\alpha_{t}= \alpha_{t-2}$, which proves conclusion
1.

Now define a trivial algorithm, called SOR2, where each iteration of
SOR2 includes two iterations of SOR. From equation (\ref{equ:sor}), we have
\begin{equation}
\label{equ:sor2}
\hat{\theta}-\tilde{\theta}_{t+1}=[I_2-(1+\alpha_{t})I_{com}^{-1}I_{obs}][I_2-(1+\alpha_{t-1})I_{com}^{-1}I_{obs}](\hat{\theta}-\tilde{\theta}_{t-1}).
\end{equation}
By conclusion 1,
$[I_p-(1+\alpha_{t})I_{com}^{-1}I_{obs}][I_p-(1+\alpha_{t-1})I_{com}^{-1}I_{obs}]$
is a constant matrix and denote it by $DM^{SOR2}$, which obviously determines the convergence rate of SOR2.
By using equation (\ref{equ:ICIOdecom}), we have
$DM^{SOR2}=I_{com}^{-1/2}T[I_p-(1+\alpha_{t})\Lambda][I_p-(1+\alpha_{t-1})\Lambda]
T'I_{com}^{1/2}$.
Moreover, with equation (\ref{equ:eta_2}) and (\ref{equ:alpha_sor_3}), it is easy to show that
 \begin{equation}
\label{equ:rate_1}
[1-(1+\alpha_{t})\lambda_j][1-(1+\alpha_{t-1})\lambda_j]=\frac{(\lambda_2-\lambda_1)^2}
{\lambda_1^2+\lambda_2^2+\lambda_1\lambda_2\left(\frac{\lambda_1^2}{\lambda_2^2}\frac{\tilde{\eta}_{t-1,1}^{2}}{\tilde{\eta}_{t-1,2}^2}+\frac{\lambda_2^2}{\lambda_1^2}\frac{\tilde{\eta}_{t-1,2}^{2}}{\tilde{\eta}_{t-1,1}^2}\right)},\ j=1,2.
\end{equation}
It follows that $DM^{SOR2}=[1-(1+\alpha_{t})\lambda_1][1-(1+\alpha_{t-1})\lambda_1]I_2$, which means SOR2 converges with the same rate $[1-(1+\alpha_{t})\lambda_1][1-(1+\alpha_{t-1})\lambda_1]$ along any direction. From equation (\ref{equ:rate_1}), it is easy to see that
$$[1-(1+\alpha_{t})\lambda_1][1-(1+\alpha_{t-1})\lambda_1]\leq
\frac{(\lambda_1-\lambda_2)^2}{(\lambda_1+\lambda_2)^2} =
(1-\frac{2\lambda_2}{\lambda_1+\lambda_2})^2 < (1-\lambda_2)^2.
$$
Note that $(\lambda_1-\lambda_2)/(\lambda_1+\lambda_2)$ is
the optimal convergence rate of SORF and that $1-\lambda_2$ is the
convergence rate of EM. Hence conclusion 2 follows.

Since $\lambda_1>\lambda_2$, equation (\ref{equ:rate}) implies that
$1-(1+\alpha_{t})\lambda_1<0$ and $1-(1+\alpha_{t})\lambda_2>0$. So from equation (\ref{equ:eta}), we have
$\tilde{\eta}_{t,1} \tilde{\eta}_{t-1,1}<0$ and $\tilde{\eta}_{t,2}
\tilde{\eta}_{t-1,2}>0$. This
proves the first statement in conclusion 3. Note that $\tilde{\theta}_{t+1}-\tilde{\theta}_{t}=(I-DM^{SOR2})(\hat{\theta}-\tilde{\theta}_{t-1})\propto \hat{\theta}-\tilde{\theta}_{t-1}$. Hence
$\tilde{\theta}_{t+1}-\tilde{\theta}_{t-1}$ is parallel to
$\hat{\theta}-\tilde{\theta}_{t-1}$, which concludes the second statement in conclusion 3. {\hfill} $\Box$

\section{The Convergence of DECME\_v1: Proof of Theorem
\ref{thm:AEM:DECME}}\label{app:AEM:DECME}
We prove this by induction. This version of proof is similar to the
proof of the PARTAN theorem in \citet[pp.~255-256]{luenberger:2003}.
 However, the difference between a generalised gradient direction and the gradient direction should be
 taken into account.

It is certainly true for $t=1$ since the first iteration is a line
search along the EM direction for both DECME\_v1 and AEM.

Now suppose that $\tilde{\theta}_{0}, \tilde{\theta}_{1}, \cdots,
\tilde{\theta}_{t-1}$ have been generated by AEM and
$\tilde{\theta}_{t}$ is determined by DECME\_v1. We want to show
that $\tilde{\theta}_{t}$ is the same point as that generated by
another iteration of AEM. For this to be true $\tilde{\theta}_{t}$
must be the point that maximises $L(\theta|Y_{obs})$ over the
two-dimensional plane $\tilde{\theta}_{t-1}+ \{\tilde{\theta}_{t-1}-
\tilde{\theta}_{t-2}, \theta_{t}-\tilde{\theta}_{t-1}\}$. Since we
assume that $L(\theta|Y_{obs})$ is a quadratic function with a positive
definite Hessian matrix, $L(\theta|Y_{obs})$ is strictly convex and we
only need to prove $\tilde{g}_{t}$ (gradient of $L(\theta|Y_{obs})$ at
$\tilde{\theta}_{t}$) is orthogonal to $\tilde{\theta}_{t-1}-
\tilde{\theta}_{t-2}$ and $\theta_{t}-\tilde{\theta}_{t-1}$, or
equivalently $\tilde{\theta}^{SOR}_{t}- \tilde{\theta}_{t-2}$ and
$\theta_{t}-\tilde{\theta}_{t-1}$. Since $\tilde{\theta}_{t}$
maximises $L(\theta|Y_{obs})$ along $\tilde{\theta}^{SOR}_{t}-
\tilde{\theta}_{t-2}$, $\tilde{g}_{t}$ is orthogonal to
$\tilde{\theta}^{SOR}_{t}- \tilde{\theta}_{t-2}$. Similarly,
$\tilde{g}^{SOR}_{t}$ is orthogonal to $\theta_{t}-
\tilde{\theta}_{t-1}$. Furthermore, we have
$\tilde{g}'_{t-2}(\theta_{t}-\tilde{\theta}_{t-1})=(\hat{\theta}-\tilde{\theta}_{t-2})'I_{obs}I_{com}^{-1}I_{obs}(\hat{\theta}-\tilde{\theta}_{t-1})
=(\theta_{t-1}-\tilde{\theta}_{t-2})'\tilde{g}_{t-1}=0$, where the last
identity is true due to the Expanding Subspace Theorem
\citep[p.~241]{luenberger:2003} for the conjugate direction methods.
Then $\tilde{g}'_{t}(\theta^{SOR}_{t}-\tilde{\theta}_{t-1})
=(\hat{\theta}-\tilde{\theta}_{t})'I_{obs}(\theta^{SOR}_{t}-\tilde{\theta}_{t-1})
=[\hat{\theta}-\tilde{\theta}_{t-2}-(1+\alpha_{t}^{(2)})(\tilde{\theta}^{SOR}_{t}-\tilde{\theta}_{t-2})]'I_{obs}(\theta^{SOR}_{t}-\tilde{\theta}_{t-1})
=[-\alpha_{t}^{(2)}(\hat{\theta}-\tilde{\theta}_{t-2})I_{obs}+(1+\alpha_{t}^{(2)})(\hat{\theta}-\tilde{\theta}^{SOR}_{t})'I_{obs}]'(\theta^{SOR}_{t}-\tilde{\theta}_{t-1})
=[-\alpha_{t}^{(2)}\tilde{g}_{t-2}+(1+\alpha_{t}^{(2)})\tilde{g}^{SOR}_{t}]'(\theta^{SOR}_{t}-\tilde{\theta}_{t-1})
=0$. It follows that $\tilde{g}_{t}$ is orthogonal to
$\theta^{SOR}_{t}-\tilde{\theta}_{t-1}$. \hfill $\Box$

\section{Implementation of Line Search}\label{app:line_search}
In practise, it is neither computationally feasible nor necessary to
conduct exact line search. In fact we can often achieve higher
efficiency by sacrificing accuracy in the line search routine,
although the number of iterations may increase. There are various
criterions for terminating the line search routine for a desirable
trade-off \cite[pp.~211-214]{luenberger:2003} and different
approaches have been proposed for efficient implementation of those
criterions \citep{More:Thuente:1994}. Furthermore, some
transformations to transform constrained problems into unconstrained
problems can be useful, as discussed in
\citet{Salakhutdinov03adaptiveoverrelaxed}.

Here we take a different approach to implement the line search by
taking the advantage of the fact that sometimes it is easy to figure
out the feasible region of a constrained problem along a single
line. After the feasible region is computed, many commonly used line
search routines available in standard software can be easily
applied. In our case the line search is conducted with the {\it
optimize} function in R by passing the computed interval
to the {\it optimize} function through its option {\it interval =}.
Note that the interval computed in this way is usually very wide and
some other information may be used to narrow it down for higher
efficiency. For example, we can start the line search by forcing
$\alpha>0$ for SOR. Furthermore, we control the accuracy of the line
search by setting $tol=0.01$ in the {\it optimize} function. This
choice is somehow arbitrary. One advantage is that the line search
is forced to be more accurate when the algorithm approaches the MLE
since the magnitude of the differences between consecutive estimates
usually becomes smaller with the progress of the algorithm.

\begin{comment} $\alpha^{(2)}>1$ for DECME\_v1, and $\alpha>1$
for DECME\_v2.\end{comment}

For the constraints involved in the examples, we summarise the
methods to obtain the feasible region as follows. Denote the
current estimation by $\theta$ and the search direction by $d$. Our
goal is to find the feasible region of $\alpha$ (an interval
including 0 for the examples used in this paper) for a univariate
function $f(\theta+\alpha d)$. If there are several sets of
constraints for one model, we can determine the feasible region
induced by each of them and then take their intersection. Without
loss of generality, we assume in the following that $d$ is the
counterpart of the discussed parameters in the vector representing
the search direction.

\newcounter{line_search}
\begin{list}
{\arabic{line_search}.)}{\usecounter{line_search}\itemsep=0cm}
\item The degree of freedom $\nu$ in the $t$ distribution. It is
easy to compute the boundary for $\alpha$ such that $\nu+\alpha
d>0$.

\item The mixing coefficients, $\pi_i,\ i=1,\cdots, K$, in the mixture model. There are
two types of constraints here, {\it i.e.,} $\sum_{i=1}^K \pi_i =1$
and $\pi_i\geq 0$. By using the first constraint, we only need to
consider the first $K-1$ coefficients with constraints
$\sum_{i=1}^{K-1} \pi_i \leq 1$ and $\pi_i \geq 0, \ i=1,\cdots,
K-1$. Then we only need to find the intersection of the solutions
for the inequalities $\sum_{i=1}^{K-1} \pi_i + \alpha
\sum_{i=1}^{K-1} d_i \leq 1$ and $\pi_i + \alpha d_i \geq 0, \
i=1,\cdots, K-1$.

\item The variance components in the linear mixed-effects model and the mixture model and the uniquenesses in the factor analysis model. This can be handled in the
same way as that for the degree of freedom in the $t$ distribution.

\item The covariance matrices in the linear mixed-effects model and the
$t$ distribution. For the current paper, only two-dimensional
covariance matrices are involved. A two-dimensional matrix $\Psi$ is
positive definite if and only if $\Psi_{1,1}>0$ and $\det(\Psi)>0$.
Hence we only need to guarantee $\Psi_{1,1} + \alpha d_{1,1} >0$ and
 $\det(\Psi+\alpha D)>0$ (assume $D$ is the matrix generated from the vector $d$ in the same way as
$\Psi$ is generated from $\theta$). For other covariance matrices of
fairly small
 size, similar method could be used. When the dimension of the covariance matrix is high,
it is a common practise to enforce certain structure on the matrix.
For example, in spatial statistics, the covariance
 matrices are usually assumed to be generated from various covariance
 functions with very few parameters  \citep{Zhang:2002, Zhu:Eickhoff:Yan:2005,
Zhang:2007} and the feasible region of $\alpha$ can be easily
obtained.
\end{list}

\bibliography{DECME}

\begin{thebibliography}{}

\bibitem[\protect\citeauthoryear{Bishop}{Bishop}{2006}]{Bishop:2006}
Bishop, C.~M. (2006).
\newblock {\em Pattern recognition and machine learning}.
\newblock Information Science and Statistics. New York: Springer.

\bibitem[\protect\citeauthoryear{Celeux, Chr{\'e}tien, Forbes, and
  Mkhadri}{Celeux et~al.}{2001}]{Celeux:Chretien:Forbes:2001}
Celeux, G., S.~Chr{\'e}tien, F.~Forbes, and A.~Mkhadri (2001).
\newblock A component-wise {EM} algorithm for mixtures.
\newblock {\em J. Comput. Graph. Statist.\/}~{\em 10\/}(4), 697--712.

\bibitem[\protect\citeauthoryear{Cleveland}{Cleveland}{1979}]{Cleveland:1979}
Cleveland, W.~S. (1979).
\newblock Robust locally weighted regression and smoothing scatterplots.
\newblock {\em Journal of the American Statistical Association\/}~{\em 74},
  829--836.

\bibitem[\protect\citeauthoryear{Cohen, Dalal, and Tukey}{Cohen
  et~al.}{1993}]{Cohen:Dalal:Tukey:robu:1993}
Cohen, M., S.~R. Dalal, and J.~W. Tukey (1993).
\newblock Robust, smoothly heterogeneous variance regression.
\newblock {\em Journal of the Royal Statistical Society, Series C: Applied
  Statistics\/}~{\em 42}, 339--353.

\bibitem[\protect\citeauthoryear{Concus, Golub, and O'Leary}{Concus
  et~al.}{1976}]{Concus:Golub:O'Leary:1976}
Concus, P., G.~H. Golub, and D.~P. O'Leary (1976).
\newblock A generalized conjugate gradient method for the numerical solution of
  elliptic partial differential equations.
\newblock In {\em Sparse matrix computations ({P}roc. {S}ympos., {A}rgonne
  {N}at. {L}ab., {L}emont, {I}ll., 1975)}, pp.\  309--332. New York: Academic
  Press.

\bibitem[\protect\citeauthoryear{Dasgupta and Schulman}{Dasgupta and
  Schulman}{2000}]{Dasgupta:Schulman:2000}
Dasgupta, S. and L.~J. Schulman (2000).
\newblock A two-round variant of em for gaussian mixtures.
\newblock In {\em UAI '00: Proceedings of the 16th Conference on Uncertainty in
  Artificial Intelligence}, San Francisco, CA, USA, pp.\  152--159. Morgan
  Kaufmann Publishers Inc.

\bibitem[\protect\citeauthoryear{Dempster, Laird, and Rubin}{Dempster
  et~al.}{1977}]{DLR:1977}
Dempster, A.~P., N.~M. Laird, and D.~B. Rubin (1977).
\newblock Maximum likelihood from incomplete data via the {EM} algorithm.
\newblock {\em Journal of the Royal Statistical Society, Series B:
  Methodological\/}~{\em 39}, 1--22.

\bibitem[\protect\citeauthoryear{Fletcher and Reeves}{Fletcher and
  Reeves}{1964}]{Fletcher:Reeves:1964}
Fletcher, R. and C.~M. Reeves (1964).
\newblock Function minimization by conjugate gradients.
\newblock {\em Comput. J.\/}~{\em 7}, 149--154.

\bibitem[\protect\citeauthoryear{Frankel}{Frankel}{1950}]{frankel:1950}
Frankel, S. (1950).
\newblock {Convergence rates of iterative treatments of partial differential
  equations}.
\newblock {\em Mathematical Tables and Other Aids to Computation\/}~{\em
  4\/}(30), 65--75.

\bibitem[\protect\citeauthoryear{Gelfand, Hills, Racine-Poon, and
  Smith}{Gelfand et~al.}{1990}]{Gelfand:Hills:1990}
Gelfand, A.~E., S.~E. Hills, A.~Racine-Poon, and A.~F.~M. Smith (1990).
\newblock Illustration of {B}ayesian inference in normal data models using
  {G}ibbs sampling.
\newblock {\em Journal of the American Statistical Association\/}~{\em 85},
  972--985.

\bibitem[\protect\citeauthoryear{Golub and Nash}{Golub and
  Nash}{1982}]{Golub:Nash:1982}
Golub, G.~H. and S.~G. Nash (1982).
\newblock Nonorthogonal analysis of variance using a generalized
  conjugate-gradient algorithm.
\newblock {\em J. Amer. Statist. Assoc.\/}~{\em 77\/}(377), 109--116.

\bibitem[\protect\citeauthoryear{Hestenes and Stiefel}{Hestenes and
  Stiefel}{1952}]{Hestenes:Stiefel:1952}
Hestenes, M.~R. and E.~Stiefel (1952).
\newblock Methods of conjugate gradients for solving linear systems.
\newblock {\em J. Research Nat. Bur. Standards\/}~{\em 49}, 409--436 (1953).

\bibitem[\protect\citeauthoryear{Hesterberg}{Hesterberg}{2005}]{Hesterberg:sta%
g:2005}
Hesterberg, T. (2005).
\newblock Staggered {A}itken acceleration for {EM}.
\newblock In {\em ASA Proceedings of the Joint Statistical Meetings}, pp.\
  2101--2110. American Statistical Association.

\bibitem[\protect\citeauthoryear{Hunter and Lange}{Hunter and
  Lange}{2004}]{hunter:lange:2004}
Hunter, D. and K.~Lange (2004).
\newblock {A Tutorial on MM Algorithms.}
\newblock {\em The American Statistician\/}~{\em 58\/}(1), 30--38.

\bibitem[\protect\citeauthoryear{Jamshidian and Jennrich}{Jamshidian and
  Jennrich}{1993}]{Jamshidian:1993}
Jamshidian, M. and R.~I. Jennrich (1993).
\newblock Conjugate gradient acceleration of the em algorithm.
\newblock {\em Journal of the American Statistical Association\/}~{\em
  88\/}(421), 221--228.

\bibitem[\protect\citeauthoryear{Jordan and Jacobs}{Jordan and
  Jacobs}{1994}]{Jordan:Robert:1994}
Jordan, M.~I. and R.~A. Jacobs (1994).
\newblock Hierarchical mixtures of experts and the em algorithm.
\newblock {\em Neural Computation\/}~{\em 6}, 181--214.

\bibitem[\protect\citeauthoryear{J\"{o}reskog}{J\"{o}reskog}{1969}]{joreskog:1%
969}
J\"{o}reskog, K. (1969, June).
\newblock A general approach to confirmatory maximum likelihood factor
  analysis.
\newblock {\em Psychometrika\/}~{\em 34\/}(2), 183--202.

\bibitem[\protect\citeauthoryear{Kowalski, Tu, Day, and
  Mendoza-Blanco}{Kowalski et~al.}{1997}]{kowalski:Tu:Day:1997}
Kowalski, J., X.~Tu, R.~Day, and J.~Mendoza-Blanco (1997).
\newblock {On the rate of convergence of the ECME algorithm for multiple
  regression models with t-distributed errors}.
\newblock {\em Biometrika\/}~{\em 84\/}(2), 269.

\bibitem[\protect\citeauthoryear{Laird and Ware}{Laird and
  Ware}{1982}]{Laird:Ware:1982}
Laird, N.~M. and J.~H. Ware (1982).
\newblock Random-effects models for longitudinal data.
\newblock {\em Biometrics\/}~{\em 38}, 963--974.

\bibitem[\protect\citeauthoryear{Lange}{Lange}{1995}]{Lang:1995b}
Lange, K. (1995).
\newblock A gradient algorithm locally equivalent to the {EM} algorithm.
\newblock {\em J. Roy. Statist. Soc. Ser. B\/}~{\em 57\/}(2), 425--437.

\bibitem[\protect\citeauthoryear{Liu}{Liu}{1998}]{Liu:infoM:1998}
Liu, C. (1998).
\newblock Information matrix computation from conditional information via
  normal approximation.
\newblock {\em Biometrika\/}~{\em 85}, 973--979.

\bibitem[\protect\citeauthoryear{Liu and Rubin}{Liu and
  Rubin}{1994}]{Liu:Rubin:ecme:1994}
Liu, C. and D.~B. Rubin (1994).
\newblock The {ECME} algorithm: {A} simple extension of {EM} and {ECM} with
  faster monotone convergence.
\newblock {\em Biometrika\/}~{\em 81}, 633--648.

\bibitem[\protect\citeauthoryear{Liu and Rubin}{Liu and
  Rubin}{1998}]{Liu:Rubin:fa:1998}
Liu, C. and D.~B. Rubin (1998).
\newblock Maximum likelihood estimation of factor analysis using the {ECME}
  algorithm with complete and incomplete data.
\newblock {\em Statist. Sinica\/}~{\em 8\/}(3), 729--747.

\bibitem[\protect\citeauthoryear{Liu and Sun}{Liu and
  Sun}{1997}]{Liu:Sun:ecme:mixture:1997}
Liu, C. and D.~X. Sun (1997).
\newblock Acceleration of {EM} algorithm for mixture models using {ECME}.
\newblock In {\em ASA Proceedings of the Statistical Computing Section}, pp.\
  109--114. American Statistical Association.

\bibitem[\protect\citeauthoryear{Luenberger}{Luenberger}{2003}]{luenberger:200%
3}
Luenberger, D. (2003).
\newblock {\em Linear and Nonlinear Programming\/} (2nd ed.).
\newblock Springer.

\bibitem[\protect\citeauthoryear{McLachlan and Peel}{McLachlan and
  Peel}{2000}]{McLachlan:Peel:2000}
McLachlan, G. and D.~Peel (2000).
\newblock {\em Finite mixture models}.
\newblock Wiley Series in Probability and Statistics: Applied Probability and
  Statistics. Wiley-Interscience, New York.

\bibitem[\protect\citeauthoryear{McLachlan and Krishnan}{McLachlan and
  Krishnan}{1997}]{McLachlan:Krishnan:1997}
McLachlan, G.~J. and T.~Krishnan (1997).
\newblock {\em The {EM} algorithm and extensions}.
\newblock Wiley Series in Probability and Statistics: Applied Probability and
  Statistics. New York: John Wiley \& Sons Inc.
\newblock A Wiley-Interscience Publication.

\bibitem[\protect\citeauthoryear{Meng and Rubin}{Meng and
  Rubin}{1993}]{Meng:Rubin:ECM:1993}
Meng, X. and D.~B. Rubin (1993).
\newblock Maximum likelihood estimation via the {ECM} algorithm: {A} general
  framework.
\newblock {\em Biometrika\/}~{\em 80}, 267--278.

\bibitem[\protect\citeauthoryear{Meng and Rubin}{Meng and
  Rubin}{1994}]{Meng:Rubin:convergenceEM:1994}
Meng, X. and D.~B. Rubin (1994).
\newblock On the global and componentwise rates of convergence of the {EM}
  algorithm ({STMA} {V}36 1300).
\newblock {\em Linear Algebra and its Applications\/}~{\em 199}, 413--425.

\bibitem[\protect\citeauthoryear{Meng and van Dyk}{Meng and van
  Dyk}{1997}]{Meng:van:AEM:1997}
Meng, X. and D.~van Dyk (1997).
\newblock The {EM} algorithm -- {A}n old folk-song sung to a fast new tune
  ({D}isc: P541-567).
\newblock {\em Journal of the Royal Statistical Society, Series B:
  Methodological\/}~{\em 59}, 511--540.

\bibitem[\protect\citeauthoryear{Mor{\'e} and Thuente}{Mor{\'e} and
  Thuente}{1994}]{More:Thuente:1994}
Mor{\'e}, J.~J. and D.~J. Thuente (1994).
\newblock Line search algorithms with guaranteed sufficient decrease.
\newblock {\em ACM Trans. Math. Software\/}~{\em 20\/}(3), 286--307.

\bibitem[\protect\citeauthoryear{Pilla and Lindsay}{Pilla and
  Lindsay}{2001}]{Pilla:Lindsay:2001}
Pilla, R.~S. and B.~G. Lindsay (2001).
\newblock Alternative {EM} methods for nonparametric finite mixture models.
\newblock {\em Biometrika\/}~{\em 88\/}(2), 535--550.

\bibitem[\protect\citeauthoryear{Pinheiro, Liu, and Wu}{Pinheiro
  et~al.}{2001}]{Pinheiro:Liu:Wu:2001}
Pinheiro, J.~C., C.~Liu, and Y.~Wu (2001).
\newblock {Efficient algorithms for robust estimation in linear mixed-effects
  models using the multivariate t distribution}.
\newblock {\em Journal of Computational and Graphical Statistics\/}~{\em
  10\/}(2), 249--276.

\bibitem[\protect\citeauthoryear{{R Development Core Team}}{{R Development Core
  Team}}{2008}]{R}
{R Development Core Team} (2008).
\newblock {\em R: A Language and Environment for Statistical Computing}.
\newblock Vienna, Austria: R Foundation for Statistical Computing.
\newblock {ISBN} 3-900051-07-0.

\bibitem[\protect\citeauthoryear{Redner and Walker}{Redner and
  Walker}{1984}]{Redner:Walker:1984}
Redner, R.~A. and H.~F. Walker (1984).
\newblock Mixture densities, maximum likelihood and the {EM} algorithm.
\newblock {\em SIAM Review\/}~{\em 26}, 195--202.

\bibitem[\protect\citeauthoryear{Rubin and Thayer}{Rubin and
  Thayer}{1982}]{Rubin:Thayer:emFA:1982}
Rubin, D.~B. and D.~T. Thayer (1982).
\newblock {EM} algorithms for {ML} factor analysis.
\newblock {\em Psychometrika\/}~{\em 47}, 69--76.

\bibitem[\protect\citeauthoryear{Salakhutdinov and Roweis}{Salakhutdinov and
  Roweis}{2003}]{Salakhutdinov03adaptiveoverrelaxed}
Salakhutdinov, R. and S.~Roweis (2003).
\newblock Adaptive overrelaxed bound optimization methods.
\newblock In {\em In Proceedings of International Conference on Machine
  Learning, ICML. International Conference on Machine Learning, ICML}, pp.\
  664--671.

\bibitem[\protect\citeauthoryear{Sammel and Ryan}{Sammel and
  Ryan}{1996}]{sammel:Ryan:1996}
Sammel, M. and L.~Ryan (1996).
\newblock {Latent variable models with fixed effects}.
\newblock {\em Biometrics\/}~{\em 52\/}(2), 650--663.

\bibitem[\protect\citeauthoryear{Shah, Buehler, and Kempthorne}{Shah
  et~al.}{1964}]{Shah:Buehler:Kempthorne:1964}
Shah, B.~V., R.~J. Buehler, and O.~Kempthorne (1964).
\newblock Some algorithms for minimizing a function of several variables.
\newblock {\em J. Soc. Indust. Appl. Math.\/}~{\em 12}, 74--92.

\bibitem[\protect\citeauthoryear{Varadhan and Roland}{Varadhan and
  Roland}{2008}]{Varadhan:Roland:2008}
Varadhan, R. and C.~Roland (2008).
\newblock Simple and globally convergent methods for accelerating the
  convergence of any {EM} algorithm.
\newblock {\em Scand. J. Statist.\/}~{\em 35\/}(2), 335--353.

\bibitem[\protect\citeauthoryear{Wu}{Wu}{1983}]{Wu:1983}
Wu, C. F.~J. (1983).
\newblock On the convergence properties of the {EM} algorithm.
\newblock {\em The Annals of Statistics\/}~{\em 11}, 95--103.

\bibitem[\protect\citeauthoryear{Young}{Young}{1954}]{young:1954}
Young, D. (1954).
\newblock {Iterative methods for solving partial difference equations of
  elliptic type}.
\newblock {\em Transactions of the American Mathematical Society\/}~{\em
  76\/}(1), 92--111.

\bibitem[\protect\citeauthoryear{Zhang}{Zhang}{2002}]{Zhang:2002}
Zhang, H. (2002).
\newblock On estimation and prediction for spatial generalized linear mixed
  models.
\newblock {\em Biometrics\/}~{\em 58\/}(1), 129--136.

\bibitem[\protect\citeauthoryear{Zhang}{Zhang}{2007}]{Zhang:2007}
Zhang, H. (2007).
\newblock Maximum-likelihood estimation for multivariate spatial linear
  coregionalization models.
\newblock {\em EnvironMetrics\/}~{\em 18\/}(2), 125--139.

\bibitem[\protect\citeauthoryear{Zhu, Eickhoff, and Yan}{Zhu
  et~al.}{2005}]{Zhu:Eickhoff:Yan:2005}
Zhu, J., J.~C. Eickhoff, and P.~Yan (2005).
\newblock Generalized linear latent variable models for repeated measures of
  spatially correlated multivariate data.
\newblock {\em Biometrics\/}~{\em 61\/}(3), 674--683.

\end{thebibliography}
\bibliographystyle{chicago}

\clearpage
 {\small \tabcolsep0.05in
\begin{table}
\caption{Eigenvalues of $DM^{EM}$ and $DM^{ECME}$ for the Linear
Mixed-effects Model Example in Section \ref{sec:rat} \label{tbl: rateigenvalue}}
% \begin{center}
 \centering
\begin{tabular}{l cc}
 \hline \hline

Algorithm && Eigenvalues of the missing information fraction \\
\cline{1-1} \cline{3-3}

 EM && 0.9860 0.9746 0.7888 0.6706 0.5176 0.3874 0.3260 0.2710
 0.0364\\

 ECME && 0.5176 0.3874 0.3260 0.2710 0.0364 0.0000 0.0000 0.0000
 0.0000\\

 \hline
\end{tabular}
%\end{center}
\end{table}
}
 {\small \tabcolsep0.05in
\begin{table}
\caption{The Four Largest Eigenvalues and the Corresponding
Eigenvectors of $DM^{EM}$ for the Linear Mixed-effects Model Example
in Section \ref{sec:rat} \label{tbl: rat}}
 \centering
 %\begin{center}
\begin{tabular}{l cc}
 \hline \hline

Eigenvalue && Corresponding eigenvector \\
\cline{1-1} \cline{3-3}

0.9860&& (0.0000 0.0000 -0.0413 -0.9991 0.0000 0.0000 0.0000 0.0000 0.0000)\\
0.9746&&(0.0413  0.9991 0.0000 0.0000 0.0000 0.0000 0.0000 0.0000 0.0000)\\
0.7888&&(0.0000 0.0000  -0.0433 0.9991 0.0000 0.0000 0.0000 0.0000 0.0000)\\
0.6706&&(-0.0433 0.9991 0.0000 0.0000 0.0000 0.0000 0.0000 0.0000 0.0000)\\
\hline
\end{tabular}
%\end{center}
\end{table}
}

 {\small \tabcolsep0.05in
\begin{table}
\caption{The Ten Leading Eigenvalues of $DM^{EM}$ and $DM^{ECME}$
for the Factor Analysis Model Example in Section \ref{sec:battery} \label{tbl: batteryeigenvalue}}
 %\begin{center}
  \centering
\begin{tabular}{l cc}
 \hline \hline

Algorithm && Ten leading eigenvalues of the missing information fraction \\
\cline{1-1} \cline{3-3}

 EM  &&  1-2E-12 0.9992 0.9651 0.9492 0.9318 0.8972 0.8699 0.8232
 0.8197 0.7876\\

ECME-1 &&  1-2E-12 0.9979 0.9509 0.9292 0.9124 0.8725 0.8480 0.8031 0.7877 0.7539\\

ECME-2 && 0.9987 0.8715 0.7321 0.6673 0.5184 0.4770 0.4496 0.3727 0.3369 0.0000\\

\hline
\end{tabular}
%\end{center}
\end{table}
}

 {\small \tabcolsep0.05in
\begin{table}
\caption{The Two Largest Eigenvalues and the Corresponding
Eigenvectors of $DM^{EM}$ for the Factor Analysis Model Example in
Section \ref{sec:battery} \label{tbl: battery}}
 %\begin{center}
  \centering
\begin{tabular}{l cc}
 \hline \hline
Eigenvalue && Corresponding eigenvector \\
\cline{1-1} \cline{3-3}
1-2E-12 &&  0.0812  0.0934 -0.4897 -0.1335 0.0684 0.0748 0.0363 -0.0864 -0.0949 \\
&& 0.0996 0.1288 0.7171 0.1962  0.1085 0.1138 0.0954 0.2099 0.2047 \\
&&  0.0000 0.0000 0.0000 0.0000 0.0000 0.0000 0.0000 0.0000 0.0000 \\
&& 0.0000 0.0000 0.0000 0.0000 0.0000 0.0000 0.0000 0.0000 0.0000\\

0.9992 && 0.0046  0.0057 -0.0047 -0.0005 -0.0047 -0.0034 -0.0038 -0.0013 0.0005\\
&&  -0.0062 -0.0071 -0.0092 -0.0064 0.0053 0.0060 0.0045 0.0049 0.0034\\
&& 0.0267 0.0441 -0.2871 0.0013 -0.0103 -0.0177 -0.0106 -0.0139 -0.0144\\
&& -0.0028 0.0079 -0.9557 0.0111 0.0013 0.0040 -0.0011 -0.0028 0.0006\\
\hline
\end{tabular}
%\end{center}
\end{table}
}

 {\small \tabcolsep0.05in
\begin{table}
\caption{Comparison of Convergence for the Three Examples in Section
\ref{sec:rat_2}, \ref{sec:battery_2}, and \ref{sec:bt} \label{tbl: iteration}}
 %\begin{center}
  \centering
\begin{tabular}{l cccccccccccc}
 \hline \hline
 && \multicolumn{11}{c}{Number of iterations}\\
 \cline{3-13}
Example && EM  && ECME && SOR && DECME\_v1 && DECME\_v2 && DECME\_v3 \\
 \cline{1-1} \cline{3-3} \cline{5-5} \cline{7-7} \cline{9-9}
 \cline{11-11} \cline{13-13}
Mixed effect: EM && 5,968  && $\diagup$ && 918 && 104 && 133 && 166 \\
Mixed effect: ECME && $\diagup$  && 20 && 15 && 9 && 13 && 15 \\
Factor analysis && 6,672 && $\diagup$ && 1,698  && 55 && 91 && 150\\
Bivariate $t$ && 293  && $\diagup$ && 96 && 32 && 48 && 63 \\

\hline
\end{tabular}
%\end{center}
\end{table}
}

 {\small \tabcolsep0.05in
\begin{table}
\caption{Comparison of Convergence for the Three Examples in Section
\ref{sec:rat_2}, \ref{sec:battery_2}, and \ref{sec:bt}, cont'd \label{tbl: CPU_time}}
 %\begin{center}
  \centering
\begin{tabular}{l cccccccccccc}
 \hline \hline
 && \multicolumn{11}{c}{CPU time (s)}\\
 \cline{3-13}
Example && EM  && ECME && SOR && DECME\_v1 && DECME\_v2 && DECME\_v3 \\
 \cline{1-1} \cline{3-3} \cline{5-5} \cline{7-7} \cline{9-9}
 \cline{11-11} \cline{13-13}
Mixed effect: EM && 518.9  && $\diagup$ && 110.9 && 15.7 && 15.6 && 19.3 \\
Mixed effect: ECME && $\diagup$  && 1.8 && 1.8 && 1.3 && 1.6 && 1.8 \\
Factor analysis && 33.4 && $\diagup$ && 21.3  && 1.2 && 1.3 && 2.0\\
Bivariate $t$ && 1.4  && $\diagup$ && 1.5 && 0.9 && 0.9 && 1.1 \\
\hline
\end{tabular}
%\end{center}
\end{table}
}

\end{document}